\documentclass[aps,pra,amsmath,reprint,floatfix]{revtex4-1}

\usepackage{microtype} 
\usepackage{amssymb}
\usepackage{graphicx}
\usepackage{bbm}
\usepackage{bm}

\usepackage{algcompatible}
\usepackage[]{algorithm2e}

\RequirePackage[
  hyperindex,colorlinks,bookmarksnumbered,
  plainpages=true,pdfstartview=FitH]{hyperref}
\hypersetup{linkcolor=blue,urlcolor=blue,citecolor=blue} 
\usepackage{hyperref}
\usepackage[all]{hypcap}

\newcommand{\ED}{.}
\newcommand{\EC}{,}

\newcommand{\ER}[1]{Eq.~(\ref{#1})}
\newcommand{\EsR}[1]{Eqs.~(\ref{#1})}
\newcommand{\ERn}[1]{(\ref{#1})}
\renewcommand{\FR}[1]{Fig.~\ref{#1}}

\newcommand{\SR}[1]{Sec.~\ref{#1}}
\newcommand{\AR}[1]{Appendix~\ref{#1}}
\newcommand{\AlR}[1]{Algorithm~\ref{#1}}

\newcommand{\N}{\mathcal{N}}
\newcommand{\p}{^{\prime}}
\newcommand{\pp}{^{\phantom\prime}}
\newcommand{\ppp}{^{\protect\phantom\prime}} 
\newcommand{\pd}{\vphantom{\dot{G}}} 
\newcommand{\pss}{^{\vphantom{(6)}}} 
\newcommand{\dif}{\textrm{d}}

\renewcommand{\vec}[1]{\bm{#1}}

\newcounter{algcounter}
\newenvironment{alg-table}{
\addtocounter{table}{-1}
\refstepcounter{algcounter}

\begin{table}}
{\end{table}}

\begin{document}

\title{Multiloop functional renormalization group for general models}
\author{Fabian B.~Kugler}
\author{Jan von Delft}
\affiliation{Physics Department, Arnold Sommerfeld Center for Theoretical Physics, and Center for NanoScience, Ludwig-Maximilians-Universit\"at M\"unchen, Theresienstr.~37, 80333 Munich, Germany}

\date{31 January 2018}

\begin{abstract}
We present multiloop flow equations in the
functional renormalization group (fRG) framework
for the four-point vertex and self-energy,
formulated for a general fermionic many-body problem.
This generalizes the previously introduced vertex flow 
[F.\ B.\ Kugler and J.\ von Delft, \href{http://link.aps.org/doi/10.1103/PhysRevLett.120.057403}{Phys. Rev. Lett. \textbf{120}, 057403 (2018)}]
and provides the necessary 
corrections to the self-energy flow
in order to complete the derivative of all diagrams 
involved in the truncated fRG flow.
Due to its iterative one-loop structure,
the multiloop flow is well suited for numerical algorithms,
enabling improvement of many fRG computations.
We demonstrate its equivalence to a solution of 
the (first-order) parquet equations
in conjunction with the Schwinger-Dyson equation
for the self-energy. 
\end{abstract}

\maketitle

\section{Introduction}
Two of the most powerful generic methods in the study of large or open
many-body systems at intermediate coupling strength
are the parquet formalism \cite{Bickers2004, Roulet1969} and
the functional renormalization group (fRG) \cite{Metzner2012, Kopietz2010}.
As is commonly known, these frameworks are intimately related. 
However, their equivalence has only recently been established
via multiloop fRG (mfRG) flow equations, introduced 
in a case study of the X-ray-edge singularity \cite{Kugler2017}.
In this paper, we consolidate this equivalence and
formulate the mfRG flow for the general many-body problem.
For this, we generalize the multiloop vertex flow from Ref.~\onlinecite{Kugler2017},
and, to ensure full inclusion of the self-energy, we present 
two multiloop corrections to the self-energy flow. 
Altogether, the mfRG flow is shown to fully generate all parquet
diagrams for the vertex and self-energy; it is thus equivalent
to solving the (first-order) parquet equations
in conjunction with the Schwinger-Dyson equation (SDE) for the self-energy.
The parquet equations (together with the SDE) provide exact, self-consistent 
equations for the four-point vertex and self-energy, 
allowing one to describe one-particle and two-particle correlations \cite{Bickers2004}.
The only input is the totally irreducible (four-point) vertex.
Approximating it by the bare interaction yields the first-order parquet equations \cite{Roulet1969}
(or parquet approximation \cite{Bickers2004}), a solution of which generates the so-called
parquet diagrams for 
the four-point vertex and self-energy. 
The functional renormalization group provides an infinite hierarchy of exact
flow equations for vertex functions, depending on an RG scale parameter $\Lambda$.
During the flow, high-energy ($ \gtrsim \Lambda $) modes are successively integrated out,
and the full solution is obtained at $\Lambda=0$, such that one is free
in the specific way the $\Lambda$ dependence (regulator) is chosen \cite{Metzner2012, Kopietz2010}.
If one restricts the fRG flow equations to the four-point vertex and self-energy,
one is left with the six-point vertex as input.
In the typical approximation, the six-point vertex is neglected, implying
that all diagrams contributing to the flow are of the parquet type \cite{Kugler2017, Kugler2017a}.
However, due to this truncation, 
the flow equations (for both self-energy and four-point vertex)
no longer form a total derivative of diagrams w.r.t.\ the flow parameter 
$\Lambda$.
This limits the predictive power of fRG and yields results
that actually depend on the choice of regulator.
The mfRG corrections to the fRG flow 
simulate the effect of six-point
vertex contributions on parquet diagrams, 
by means of an iterative multiloop construction.
They complete the derivative of diagrams 
in the flow equations of both self-energy
and four-point vertex, which are 
otherwise only partially contained.
As it achieves a full resummation of all parquet diagrams
in a numerically efficient way,
the mfRG flow allows for significant improvement
of fRG computations and overcomes weaknesses
of the formalism experienced hitherto.
The paper is organized as follows.
In \SR{sec:setup}, we give the setup with all notations,
before we recall the basics of the parquet formalism
in \SR{sec:parquet}.
In \SR{sec:mfrg}, we present the
mfRG flow equations
for the four-point vertex and self-energy.
We show that they fully generate all parquet diagrams
to arbitrary order in the interaction and
comment on computational and general properties of the flow equations.
Finally, we present our conclusions in \SR{sec:conclusion}.
\section{Setup}
\label{sec:setup}
We consider a general theory of interacting fermions,
defined by the action
\begin{align}
S & = - \sum_{x\p, x} \bar{c}_{x\p} \big[ (G^0)^{-1} \big]_{x\p, x} c_{x} 
- \tfrac{1}{4} 
\!\!\!\! \sum_{x\p,x,y\p,y} \!\!\!\!
\Gamma^0_{x\p,y\p;x,y} \bar{c}_{x\p} \bar{c}_{y\p} c_{y} c_{x}
\EC
\end{align}
with a bare propagator $G^0$ and 
a bare four-point vertex $\Gamma^0$, which is
antisymmetric in its first and last two arguments.
The index $x$ denotes all quantum numbers of the Grassmann field $c_x$.
If we choose, e.g., Matsubara frequency, momentum, and spin, with
$x = (i\omega, \vec{k}, \sigma) = (k, \sigma)$, and consider
a translationally invariant system with interaction $U_{|\vec{k}|}$,
the bare quantities read
\begin{subequations}
\begin{align}
G^0_{x\p,x} 
& 
\overset{\textrm{e.g.}}{=}
G^0_{k,\sigma} \delta_{k\p,k\pp} \delta_{\sigma\p,\sigma\pp}
\\
- \Gamma^0_{x_1\p,x_2\p;x_1\pp,x_2\pp} 
& 
\overset{\textrm{e.g.}}{=}
(
U_{|\vec{k}_1\p-\vec{k}_1\pp|} \delta_{\sigma_1\p,\sigma_1\pp} 
\delta_{\sigma_2\p,\sigma_2\pp}
\nonumber \\ & \, \ -
U_{|\vec{k}_1\p-\vec{k}_2\pp|} \delta_{\sigma_1\p,\sigma_2\pp} 
\delta_{\sigma_2\p,\sigma_1\pp}
) \, 
\delta_{k_1\p+k_2\p,k_1\pp+k_2\pp}
\ED
\label{eq:vertex_U}
\end{align}
\end{subequations}
Correlation functions of fields, corresponding to time-ordered
expectation values of operators, are given by the path integral
\begin{equation}
\langle c_{x_1} \cdots \bar{c}_{x_n} \rangle = 
\frac{1}{Z}
\int \! \mathcal{D}[\bar{c}] \mathcal{D}[c] \,
c_{x_1} \cdots \bar{c}_{x_n} e^{-S}
\EC
\end{equation}
where $Z$ ensures normalization, such that $\langle 1 \rangle=1$.
Two-point correlation functions are represented by the full propagator $G$.
Via Dyson's equation, $G$ is expressed in terms of the 
bare propagator $G^0$ and the
self-energy $\Sigma$ [cf.~\FR{fig:dyson}(a)], according to
\begin{align}
G_{x, x\p} = - \langle c_{x} \bar{c}_{x\p} \rangle
\EC \quad
G = G^0 + G^0 \cdot \Sigma \cdot G
\EC
\label{eq:dyson}
\end{align}
using the matrix product
$
( A \cdot B )_{x, x\p} = \sum_{y} A_{x, y} B_{y, x\p}
\ED
$
In a diagrammatic expansion, the
lowest-order contribution to the self-energy
is given by the diagram in \FR{fig:dyson}(b),
making use of the bare objects $G^0$, $\Gamma^0$.
For later purposes, we define
a \textit{self-energy loop} ($L$) as
\begin{align}
L(\Gamma, G)_{x\p, x}
& =
- \sum_{y\p, y} \Gamma_{x\p, y\p; x, y} G_{y, y\p}
\ED
\label{eq:sigmaloop}
\end{align}
With this, we can write the first-order contribution
from \FR{fig:dyson}(b)
generally and in the above example as
\begin{subequations}
\begin{align}
\Sigma^{1\textrm{st}}_{x\p, x} 
& 
\overset{\hphantom{\textrm{e.g.}}}{=} 
L(\Gamma^0, G^0)_{x\p, x}
\\
&
\overset{\textrm{e.g.}}{=}
\big(
U_{0} \sum_{\tilde{k},\tilde{\sigma}} G^0_{\tilde{k},\tilde{\sigma}}
- 
\sum_{\tilde{k}} U_{|\vec{k}-\tilde{\vec{k}}|}^{\phantom0} G^0_{\tilde{k},\sigma}
\big)
\delta_{k\p,k}
\delta_{\sigma\p,\sigma}
\ED
\end{align}
\end{subequations}
\begin{figure}[t]
\includegraphics[width=.48\textwidth]{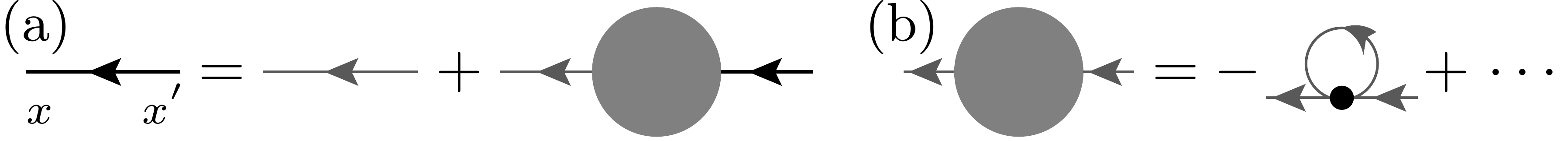}
\caption{%
(a)
Dyson's equation relating the full propagator $G_{x,x\p}$ (black, thick line)
to the bare propagator $G^0$ (gray, thin line)
and the self-energy $\Sigma$ (circle).
(b)
First-order diagram for the self-energy
using the bare vertex $\Gamma^0$ (solid dot).%
}
\label{fig:dyson}
\end{figure}
Four-point correlation functions can be expressed via the full
(one-particle-irreducible) 
four-point vertex $\Gamma$:
\begin{align}
\langle c_{x_1\pp} c_{x_2\pp} \bar{c}_{x_2\p} \bar{c}_{x_1\p} \rangle
& =
G_{x_1\pp x_1\p} G_{x_2\pp x_2\p} - G_{x_1\pp x_2\p} G_{x_2\pp x_1\p}
\nonumber \\
& \ +
G_{x_1\pp y_1\p} G_{x_2\pp y_2\p}
\Gamma_{y_1\p, y_2\p; y_1\pp, y_2\pp}
G_{y_1\pp x_1\p} G_{y_2\pp x_2\p}
\ED
\end{align}
Note that we omit the superscript compared to the
usual notation ($\Gamma^{(4)}$) \cite{Kugler2017, Kugler2017a, Metzner2012, Kopietz2010}
and often refer to 
the four-point vertex simply as the vertex.
Our definition of $\Gamma$ 
\footnote{Defining the four-point vertex as expansion coefficient of the (quantum) effective action $\vec{\Gamma}$, we  use $\Gamma_{x\p, y\p; x, y} = \delta^4 \vec{\Gamma} / ( \delta \bar{c}_{x\p} \delta \bar{c}_{y\p} \delta c_{x} \delta c_{y})$ at zero fields.
Via antisymmetry, we have $\Gamma_{x\p, y\p; x, y} = - \Gamma_{x\p, y\p; y, x}$, and all of the standard relations in our paper agree precisely with those of Ref.~\onlinecite{Kopietz2010}.}
agrees with that of Ref.~\onlinecite{Kopietz2010}
and therefore contains a relative minus sign compared to Ref.~\onlinecite{Metzner2012}.
The diagrammatic expansion of $\Gamma$ 
up to second order in the interaction
is shown in \FR{fig:vertex}.
In such diagrams,
the position of the external legs will always be fixed
and labeled in correspondence to the four arguments of a vertex.
Let us define \textit{bubble functions} ($B$),
distinguished between the three 
two-particle channels $r \in \{ a, p, t \}$, as
\begin{subequations}
\label{eq:bubbles}
\begin{align}
B_a(\Gamma, \Gamma')_{x_1\p, x_2\p; x_1\pp, x_2\pp} 
& =
\sum_{y_1\p, y_1\pp, y_2\p, y_2\pp} 
\Gamma_{x_1\p, y_2\p; y_1\pp, x_2\pp}
\nonumber \\
& \ \times
G_{y_1\pp, y_1\p} G_{y_2\pp, y_2\p} 
\Gamma'_{y_1\p, x_2\p; x_1\pp, y_2\pp}
\label{eq:abubble} \\
B_p(\Gamma, \Gamma')_{x_1\p, x_2\p; x_1\pp, x_2\pp}
& = \tfrac{1}{2}
\sum_{y_1\p, y_1\pp, y_2\p, y_2\pp} 
\Gamma_{x_1\p, x_2\p; y_1, y_2\pp}
\nonumber \\
& \ \times
G_{y_1\pp, y_1\p} G_{y_2\pp, y_2\p} 
\Gamma'_{y_1\p, y_2\p; x_1\pp, x_2\pp}
\label{eq:pbubble} \\
B_t(\Gamma, \Gamma')_{x_1\p, x_2\p; x_1\pp, x_2\pp} 
& = -
\sum_{y_1\p, y_1\pp, y_2\p, y_2\pp} 
\Gamma_{y_1\p, x_2\p; y_1\pp, x_2\pp}
\nonumber \\
& \ \times
G_{y_2\pp, y_1\p} G_{y_1\pp, y_2\p}
\Gamma'_{x_1\p, y_2\p; x_1\pp, y_2\pp}
\ED
\label{eq:tbubble}
\end{align}
\end{subequations}
The translation of \FR{fig:vertex} is then simply given by
\begin{equation}
\textstyle
\Gamma^{\textrm{2nd}}
=
\Gamma^0
+
\sum_r
B_r(\Gamma^0, \Gamma^0)
\ED
\label{eq:vertex2ndOrder}
\end{equation}
\begin{figure}[t]
\includegraphics[width=.48\textwidth]{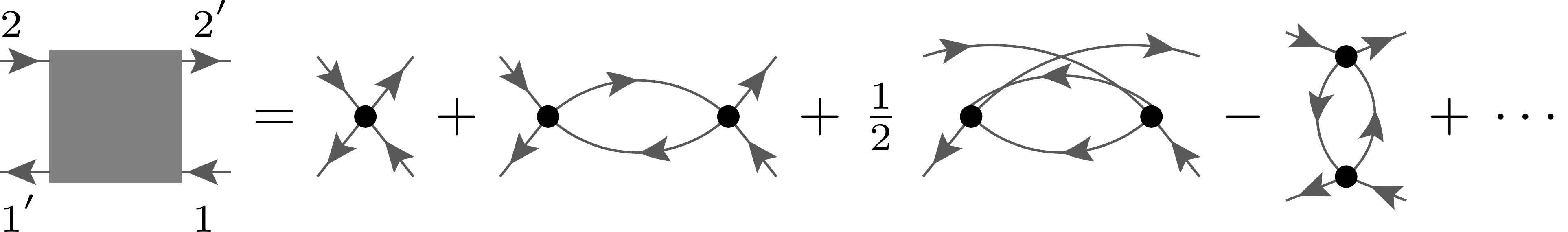}
\caption{%
Diagrammatic expansion of the four-point vertex
$\Gamma$ (square)
up to second order in the interaction
(i.e., these diagrams define $\Gamma^{\textrm{2nd}}$).
The positions of the external (amputated) legs
refer to the arguments of 
$\Gamma_{x_1\p, x_2\p; x_1\ppp, x_2\ppp}$.%
}
\label{fig:vertex}
\end{figure}
Following the conventions of Bickers \cite{Bickers2004},
the factor of $1/2$ in 
\ER{eq:pbubble} (\FR{fig:vertex})
makes sure that, when summing over all internal indices,
one does not overcount the effect of the two
indistinguishable (parallel) lines.
The minus sign in \ER{eq:tbubble} (\FR{fig:vertex})
stems from the fact that the antiparallel bubbles
\ERn{eq:abubble} and \ERn{eq:tbubble} are
related by exchange of fermionic legs. Indeed,
using the antisymmetry of $\Gamma$ and $\Gamma'$ in their arguments (crossing symmetry),
we find 
\begin{equation}
B_a(\Gamma, \Gamma')_{x_1\p, x_2\p; x_1\pp, x_2\pp} 
= -
B_t(\Gamma, \Gamma')_{x_2\p, x_1\p; x_1\pp, x_2\pp} 
\ED
\quad
\label{eq:a_t_sym}
\end{equation}
The channel label $r \in \{ a, p, t \}$ refers to the fact
that the individual diagrams are reducible---i.e.,
they fall apart into disconnected diagrams---by cutting
two \textit{antiparallel} lines,
two \textit{parallel} lines, or
two \textit{transverse} (antiparallel) lines, respectively.
(The term transverse itself refers to a horizontal space-time axis.)
In using the terms antiparallel and parallel,
we adopt the nomenclature used in the seminal application of the parquet equations
to the X-ray-edge singularity by Roulet et al.\ \cite{Roulet1969}.
Equivalently, a common notation \cite{Rohringer2017, Wentzell2016}
for the channels $a, p, t$
is $ph, pp, \overline{ph}$, referring to
the (longitudinal) particle-hole, the particle-particle,
and the transverse (or vertical) particle-hole channel, respectively.
One also finds the labels $x, p, d$ in the literature \cite{Jakobs2010}, 
referring to the
so-called exchange, pairing, and direct channel, respectively.
In the context of fRG (cf.~\SR{sec:mfrg}),
functions such as $G$, $\Sigma$, $\Gamma$ develop a 
scale ($\Lambda$) dependence (which will be suppressed in the notation).
If we write the bubble functions also symbolically as
\begin{align}
B_r(\Gamma, \Gamma') 
& =
\big[ \Gamma \circ G \circ G \circ \Gamma' \big]_r
\EC
\label{eq:gbubble}
\end{align}
we can immediately define bubbles
with differentiated propagators
(but undifferentiated vertices)
according to
\begin{align}
\dot{B}_r(\Gamma, \Gamma') 
& =
\big[ \Gamma \circ 
\big( \partial_{\Lambda} ( G \circ G ) \big) \circ \Gamma' \big]_r
\EC
\label{eq:gdotbubble}
\end{align}
In the fRG flow equations, we will further need the 
(so-called) single-scale propagator, defined by ($\mathbbm{1}_{x,y}=\delta_{x,y}$)
\begin{align}
S = 
\partial_{\Lambda} G|_{\Sigma=\textrm{const.}}
= ( \mathbbm{1} + G \cdot \Sigma ) \cdot 
\big( \partial_{\Lambda} G^0 \big) \cdot 
( \Sigma \cdot G + \mathbbm{1} )
\ED
\label{eq:singlescale}
\end{align}
Before moving on to the mfRG flow,
let us next review
the basics of the parquet formalism.
\section{Parquet formalism}
\label{sec:parquet}
The parquet formalism \cite{Bickers2004,Roulet1969}
provides exact, self-consistent equations for both
four-point vertex and self-energy.
Focusing on the vertex first, the central
parquet equation represents a classification of
diagrams distinguished by reducibility
in the three two-particle channels:
\begin{align}
\Gamma 
& = 
R + \sum_r \gamma_r
\EC \quad
I_r = R + \sum_{r' \neq r} \gamma_{r'}
\ED
\label{eq:parquet}
\end{align}
Diagrams of $\Gamma$ are either reducible in one of 
the three channels
(i.e., part of $\gamma_r$ for $r \in \{ a, p, t \}$, cf.~\FR{fig:vertex}),
or they belong to the class of 
totally irreducible diagrams $R$ [cf.~\FR{fig:sd}(a)].
(The notation again refers to Ref.~\onlinecite{Roulet1969}.) 
As a diagram cannot simultaneously 
be reducible in more than one
channel \cite{Roulet1969}, one collects diagrams that are not
reducible in $r$ lines into the irreducible vertex $I_r$
of that channel.
Reducible and irreducible vertices are
further related by the
self-consistent Bethe-Salpeter equations (BSEs)
\begin{align}
\gamma_r 
& = 
B_r(I_r, \Gamma)
\EC
\label{eq:BetheSalpeter}
\end{align}
the graphical representations of which are given in \FR{fig:bs}.
\begin{figure}[t]
\center
\includegraphics[width=.48\textwidth]{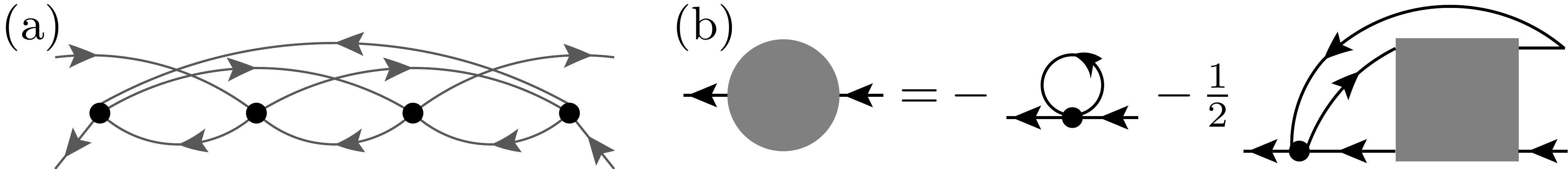}
\caption{%
(a)
Vertex diagram irreducible in all two-particle channels
(i.e., it belongs to $R$) and thus not part of $\Gamma$ in the
parquet approximation.
(b)
Schwinger-Dyson equation, relating the self-energy 
to the four-point vertex self-consistently.%
}
\label{fig:sd}
\end{figure}
The BSEs \ERn{eq:BetheSalpeter}
are computed with full propagators $G$.
Thus, they require knowledge of the self-energy,
which itself can be determined by the self-consistent 
SDE depending on the four-point vertex
[cf.~\FR{fig:sd}(b)]:
\begin{align}
\Sigma
& = 
L(\Gamma^0, G)
+ 
L\big[B_p(\Gamma^0,\Gamma),G\big]
\nonumber \\
& = 
L(\Gamma^0, G)
+ 
\tfrac{1}{2}
L\big[B_a(\Gamma^0,\Gamma),G\big]
\ED
\label{eq:SchwingerDyson}
\end{align}
%

%
\begin{figure}[t]
\includegraphics[width=.455\textwidth]{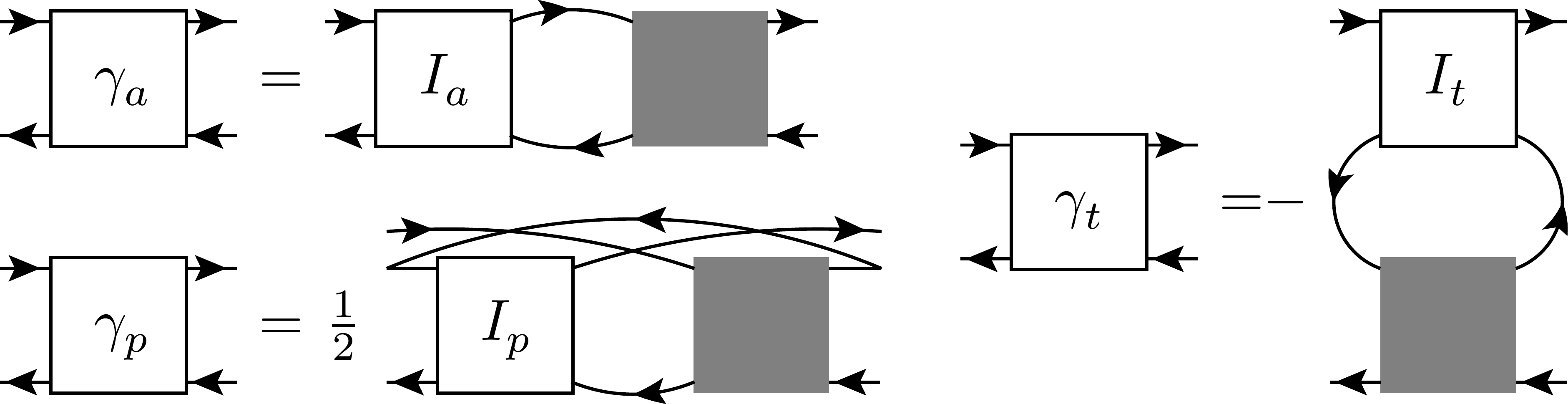}
\caption{%
Bethe-Salpeter equations in the three two-particle channels,
relating the reducible ($\gamma_r$) and irreducible ($I_r$) vertices
self-consistently in the parquet formalism.%
}
\label{fig:bs}
\end{figure}
The only input required for solving the parquet equations
is the totally irreducible vertex $R$. All remaining contributions
to the vertex and self-energy are determined self-consistently.
The simplest way to solve the parquet equations
is to approximate $R$ by the bare vertex $\Gamma^0$.
This is called the first-order parquet solution \cite{Roulet1969},
or parquet approximation \cite{Bickers2004},
and corresponds to a summation of the leading logarithmic diagrams
in logarithmically divergent perturbation theories.
The diagrams generated by the first-order
parquet solution are called parquet diagrams.
For $\Gamma$, these can be obtained by successively 
replacing bare vertices by one of the three bubbles 
from \ER{eq:bubbles}
(connected by full lines),
starting from the bare vertex.
For $\Sigma$, the parquet diagrams are obtained by
inserting the parquet vertex into the SDE. 
They can also be characterized by the property
that one needs to cut at most
one bare line to obtain a \textit{parquet} vertex
with possible dressing at the external legs.
By this, we mean that, instead of an ingoing
or outgoing amputated leg,
the external line is of the type 
$\mathbbm{1} + \Sigma \cdot G$ or
$\mathbbm{1} + G \cdot \Sigma$, respectively,
using again a parquet self-energy.
\section{Multiloop fRG flow}
\label{sec:mfrg}
The functional renormalization group \cite{Metzner2012,Kopietz2010}
provides a hierarchy of exact flow equations for
vertex functions, depending on an RG parameter
$\Lambda$, serving as infrared cutoff in the bare propagator.
A typical choice for the $\Lambda$ dependence,
in order to flow from the trivially uncorrelated
to the full theory, is characterized by the boundary conditions
$G_{\Lambda_i}=0$ and $G_{\Lambda_f}=G$,
implying $\Gamma_{\Lambda_i}=\Gamma^0$.
Restring the flow to $\Sigma$ and $\Gamma$, the six-point
vertex remains as input and is
neglected in the standard approximation.
\begin{figure*}[t]
\includegraphics[width=.99\textwidth]{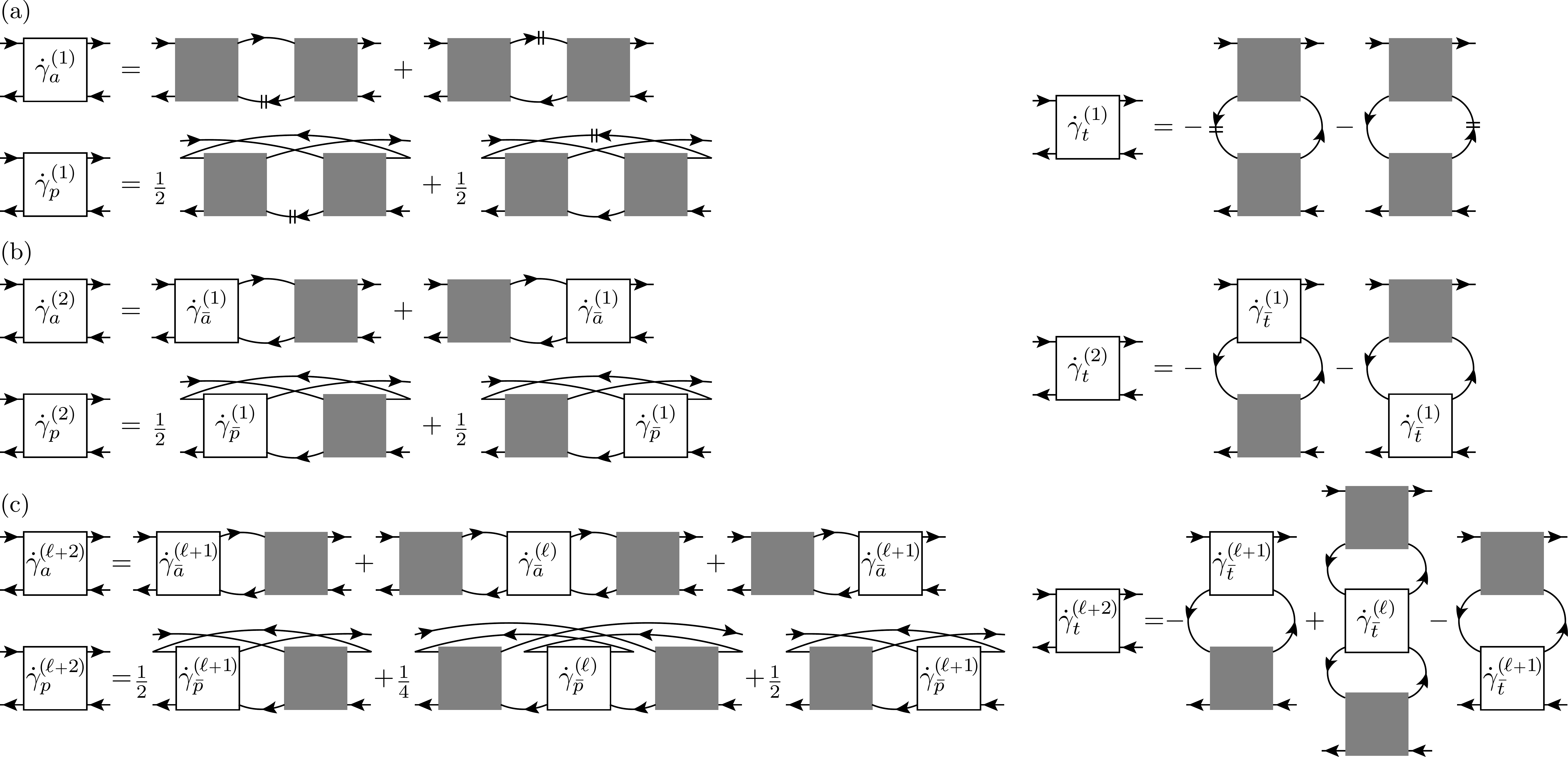}
\caption{%
Multiloop flow equations for the four-point vertex in a general fermionic model.
(a) Standard truncated, one-loop flow,
where a line with double dashes denotes $\partial_{\Lambda} G$.
(b) Two-loop correction
(upon inserting the one-loop contributions, one obtains 
two loops connecting full vertices).
(c) Higher-loop corrections starting from $\ell+2=3$,
which contain the additional contribution 
(center part) where vertices
from the complementary channels are connected by two bubbles.
}
\label{fig:vertex_flow}
\end{figure*}
Here, we view fRG as a tool to resum diagrams
which does not necessarily rely on the original 
fRG hierarchy deduced from the flow of the
(quantum) effective action.
In previous works \cite{Kugler2017, Kugler2017a}, 
we have used the X-ray-edge singularity
as an example to show that
the standard truncation of fRG
restricts the flow to parquet diagrams
of the vertex, 
and that the derivatives of those diagrams
are only partially contained.
Using the same model, we have introduced
multiloop fRG flow equations for the vertex 
which complete the derivative of parquet diagrams
in an iterative manner, as organized 
by the number of
loops connecting full vertices,
and thus do achieve a full summation
of all parquet diagrams \cite{Kugler2017}.
The X-ray-edge singularity
facilitates diagrammatic arguments as it
allows one to consider only 
two two-particle channels and 
to neglect self-energies. 
Here, we give the details of how the mfRG flow
of the vertex is generalized
to all three two-particle channels 
with indistinguishable particles 
(as already indicated in Ref.~\onlinecite{Kugler2017})
and formulate the mfRG corrections
to the self-energy flow (not discussed in Ref.~\onlinecite{Kugler2017}).
We first pose the mfRG flow equations and motivate them by
showing examples of diagrams, which are otherwise only partially contained.
Then, we justify the extensions of the truncated fRG flow by
arguing that all diagrams are of the appropriate
type without any overcounting.
Subsequently, we give a recipe for counting
the number of diagrams generated by the parquet and
mfRG flow equations. This allows one to check
that the mfRG flow fully captures all parquet diagrams 
order for order in the interaction.
Finally, we discuss computational and general properties of the flow equations.
\subsection{Flow equations for the vertex}
The mfRG flow of the vertex 
proposed in Ref.~\onlinecite{Kugler2017} makes use of the 
channel classification known from the parquet equations
and is organized by the \textit{loop order} $\ell$.
We write
\begin{align}
\partial_{\Lambda} \Gamma =
\sum_r \partial_{\Lambda} \gamma_r  
\EC \ 
\partial_{\Lambda} \gamma_r 
= 
\sum_{\ell \geq 1} \dot{\gamma}_r^{(\ell)}
\EC \ 
\dot{\gamma}_{\bar{r}}^{(\ell)} = \sum_{r' \neq r} \dot{\gamma}_{r'}^{(\ell)}
\EC
\end{align}
where $\dot{\gamma}_r^{(\ell)}$ contains
differentiated diagrams
reducible in channel $r$
with $\ell$ loops connecting full vertices
and will be constructed iteratively;
$\bar{r}$ represents the complementary channels
to channel $r$.
Using the bubble functions \ERn{eq:bubbles}
and the channel decomposition, the multiloop flow
for $\Gamma$
is compactly stated as ($\ell \geq 1$)
\begin{subequations}
\label{eq:multiloop_flow}
\begin{align}
\dot{\gamma}_r^{(1)}
& = 
\dot{B}_r(\Gamma, \Gamma)
\label{eq:one-loop_flow} \EC \\
\dot{\gamma}_r^{(2)}
& = 
B_r \big( \dot{\gamma}_{\bar{r}}^{(1)}, \Gamma \big)
+
B_r \big( \Gamma, \dot{\gamma}_{\bar{r}}^{(1)} \big) 
\label{eq:two-loop_flow} \EC \\
\dot{\gamma}_r^{(\ell+2)}
& = 
B_r \big( \dot{\gamma}_{\bar{r}}^{(\ell+1)}, \Gamma \big)
+ 
\dot{\gamma}_{r,\textrm{C}}^{(\ell+2)}
+
B_r \big( \Gamma, \dot{\gamma}_{\bar{r}}^{(\ell+1)} \big)
\label{eq:higher-loop_flow} \EC \\
\dot{\gamma}_{r,\textrm{C}}^{(\ell+2)}
& = 
B_r \big[ \Gamma, 
B_r \big( \dot{\gamma}_{\bar{r}}^{(\ell)}, \Gamma \big) \big]
= 
B_r \big[ B_r \big( \Gamma, \dot{\gamma}_{\bar{r}}^{(\ell)} \big),
\Gamma \big]
\label{eq:vertex_center_part}
\end{align}
\end{subequations}
and illustrated in \FR{fig:vertex_flow}.

The standard truncated, one-loop flow 
of $\Gamma$ is simply given by \ER{eq:one-loop_flow} [\FR{fig:vertex_flow}(a)].
A simplified version of this equation,
in which one uses the single-scale propagator $S$ \ERn{eq:singlescale}
instead of $\partial_{\Lambda}G$ 
in the differentiated bubble \ERn{eq:gdotbubble},
corresponds to the result obtained from the exact flow equation
upon neglecting the six-point vertex 
\footnote{Note that in the flow equation of the vertex in Ref.~\onlinecite{Metzner2012}, Eq.~(52), a minus sign in the first line on the r.h.s.\ is missing.}.
The form given here,
with $\partial_{\Lambda}G$ instead of $S$
(also known as Katanin substitution 
\cite{Metzner2012, Katanin2004}),
already includes corrections to this
originating from vertex diagrams 
containing differentiated
self-energy contributions.
In the exact flow equation, these contributions
are contained in the six-point vertex
$\Gamma^{(6)}$ and excluded in $S$; omitting $\Gamma^{(6)}$,
they are incorporated again by 
$\partial_{\Lambda}G = S + G \cdot (\partial_{\Lambda} \Sigma) \cdot G$.
Comparing \EsR{eq:vertex2ndOrder}, \ERn{eq:gbubble}, \ERn{eq:gdotbubble} 
with \ER{eq:one-loop_flow}
[or \FR{fig:vertex} with \FR{fig:vertex_flow}(a)],
it is clear that the one-loop flow is correct up to
second order, for which only bare vertices are involved.
Indeed, all differentiated diagrams of $\Gamma^{2^{\textrm{nd}}}$,
which are obtained by summing all copies of diagrams in which
one $G^0$ line is replaced by $\partial_{\Lambda}G^0$,
are contained in $\sum_r \dot{\gamma}_r^{(1)}$.
However, starting at third order, the one-loop flow \ERn{eq:one-loop_flow} 
does not fully generate all (parquet) diagrams, since,
in the exact flow, the six-point vertex starts contributing. 
In mfRG, the two-loop flow
[\ER{eq:two-loop_flow}, \FR{fig:vertex_flow}(b)] completes
the derivative of third-order diagrams of $\Gamma$
(i.e., it contains all diagrams needed to ensure that
$\dot{\gamma}_r^{(1)}+\dot{\gamma}_r^{(2)}$ fully represent
$\partial_{\Lambda} \gamma_r^{\textrm{3rd}}$).
An example is given in \FR{fig:examples}(a), which 
shows a parquet diagram reducible in channel $a$.
The differentiated diagram in \FR{fig:examples}(d),
as part of the derivative of \FR{fig:examples}(a),
is not included in the one-loop flow.
The reason is that $\dot{\gamma}_{a}^{(1)}$ only
contains vertices connected by antiparallel $G^0$-$\partial_{\Lambda}G^0$ lines,
and not parallel ones,
as would be necessary for this differentiated diagram.
It is, however, included in the two-loop correction to the flow,
as can be seen by inserting the lowest-order
contributions for all vertices into the first summand
on the r.h.s.\ of $\dot{\gamma}_{a}^{(2)}$ (using $\dot{\gamma}_{p}^{(1)}$) in 
\FR{fig:vertex_flow}(b).
\begin{figure}[t]
\center
\includegraphics[width=.48\textwidth]{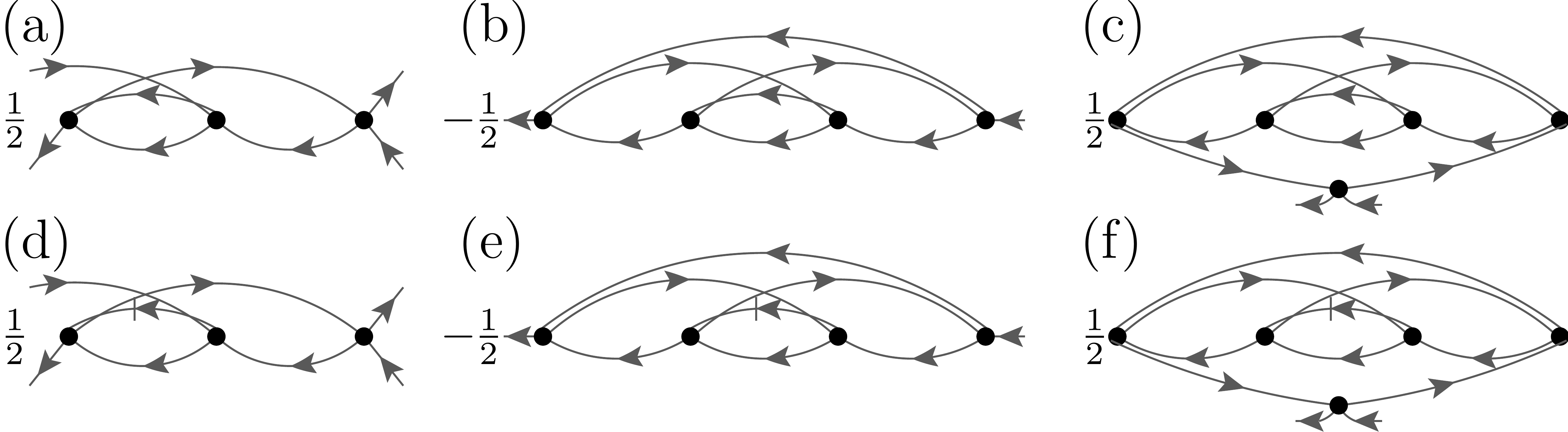}
\caption{%
(a-c) Some diagrams that are included in the
parquet approximation and only
partially contained in one-loop fRG.
(d-f) One particular differentiated diagram
for each of the diagrams (a-c) 
[the (gray, thin) line with a dash
stands for $\partial_{\Lambda} G^0$]
that is not part of the standard truncated
flow, but included in mfRG.%
}
\label{fig:examples}
\end{figure}
At all higher loop orders ($\ell+2 \geq 3$)
[\ER{eq:higher-loop_flow}, \FR{fig:vertex_flow}(c)],
we iterate this scheme and further add the 
\textit{center part} \ERn{eq:vertex_center_part} of the vertex flow.
This connects the $\ell$-loop flow 
from the complementary ($\bar{r}$) channels 
by $r$ bubbles on both sides,
and is needed to complete the derivative
of parquet diagrams starting at fourth order.
Since $\dot{\gamma}_{r,\textrm{C}}^{(\ell+2)}$ raises the loop order by two,
it was still absent in the two-loop flow. The three summands in 
$\dot{\gamma}_{r}^{(\ell+2)}$, including $\dot{\gamma}_{r,\textrm{C}}^{(\ell+2)}$,
exhaust all possibilities to obtain differentiated vertex diagrams in channel $r$ at loop order $\ell+2$
in an iterative one-loop procedure. The mfRG vertex flow up to loop order $\ell$
therefore fully captures all parquet diagrams up to order $n=\ell+1$ in the interaction
(cf.~\SR{sec:diagr_count}). 
\subsection{Flow equation for the self-energy}
The self-energy has an \textit{exact} fRG flow equation, which simply connects
the four-point vertex with the single-scale propagator (cf.~\FR{fig:sigma_flow}). 
However, if a vertex obtained from the truncated vertex flow is
inserted into this standard self-energy flow equation,
it generates diagrams that are only partially differentiated.
In fact, even after correcting the vertex flow via mfRG
to obtain all parquet diagrams of $\Gamma$, 
$\dot{\Sigma}_{\textrm{std}}$ does not yet form a total derivative.
Although $\dot{\Sigma}_{\textrm{std}}$ is in principle exact
[as is the SDE \ERn{eq:SchwingerDyson}],
using the \textit{parquet} vertex in this
flow gives a less accurate result than
inserting it into the SDE:
All diagrams obtained from $\dot{\Sigma}_{\textrm{std}}$ are of the parquet type,
but their derivatives are not fully generated by
the standard flow equation.
This problem can be remedied by adding multiloop corrections
to the self-energy flow, which complete the derivative 
of all involved diagrams. The corrections consist of two additions 
that build on the center parts \ERn{eq:vertex_center_part} of the vertex flow
in the $a$ and $p$ channels,
\begin{equation}
\dot{\gamma}_{\bar{t},\textrm{C}} = 
\sum_{\ell \geq 1}
\big(
\dot{\gamma}_{a,\textrm{C}}^{(\ell)} +
\dot{\gamma}_{p,\textrm{C}}^{(\ell)} \big)
\ED
\label{eq:gamma_tbar_c}
\end{equation}
Using the self-energy loop 
\ERn{eq:sigmaloop},
the mfRG flow equation for $\Sigma$ is then given by
(cf.\ \FR{fig:sigma_flow})
\begin{subequations}
\label{eq:sigma_flow}
\begin{flalign}
\partial_{\Lambda} \Sigma
& = 
\dot{\Sigma}_{\textrm{std}}
+ \dot{\Sigma}_{\bar{t}} + \dot{\Sigma}_{t}
\EC
&
\dot{\Sigma}_{\textrm{std}}
& = 
L(\Gamma, S)
\EC
\label{eq:sigma_flow_std} \\
\dot{\Sigma}_{\bar{t}}
& = 
L(\dot{\gamma}_{\bar{t},\textrm{C}}, G)
\EC
&
\dot{\Sigma}_{t}
& =
L ( \Gamma, G \cdot \dot{\Sigma}_{\bar{t}} \cdot G )
\ED
\label{eq:sigma_flow_t}
\end{flalign}
\end{subequations}
Note that self-energy diagrams in $\dot{\Sigma}_{t}$ and
$\dot{\Sigma}_{\bar{t}}$ are reducible and irreducible in the $t$ channel,
respectively.
However, here, this property is not exclusive;
$\dot{\Sigma}_{\textrm{std}}$, too, contains diagrams that are
reducible and irreducible in the $t$ channel, as is directly seen by inserting
the second-order vertex from \FR{fig:vertex} into the first summand of \FR{fig:sigma_flow}.
\begin{figure}[t]
\center
\includegraphics[width=.4\textwidth]{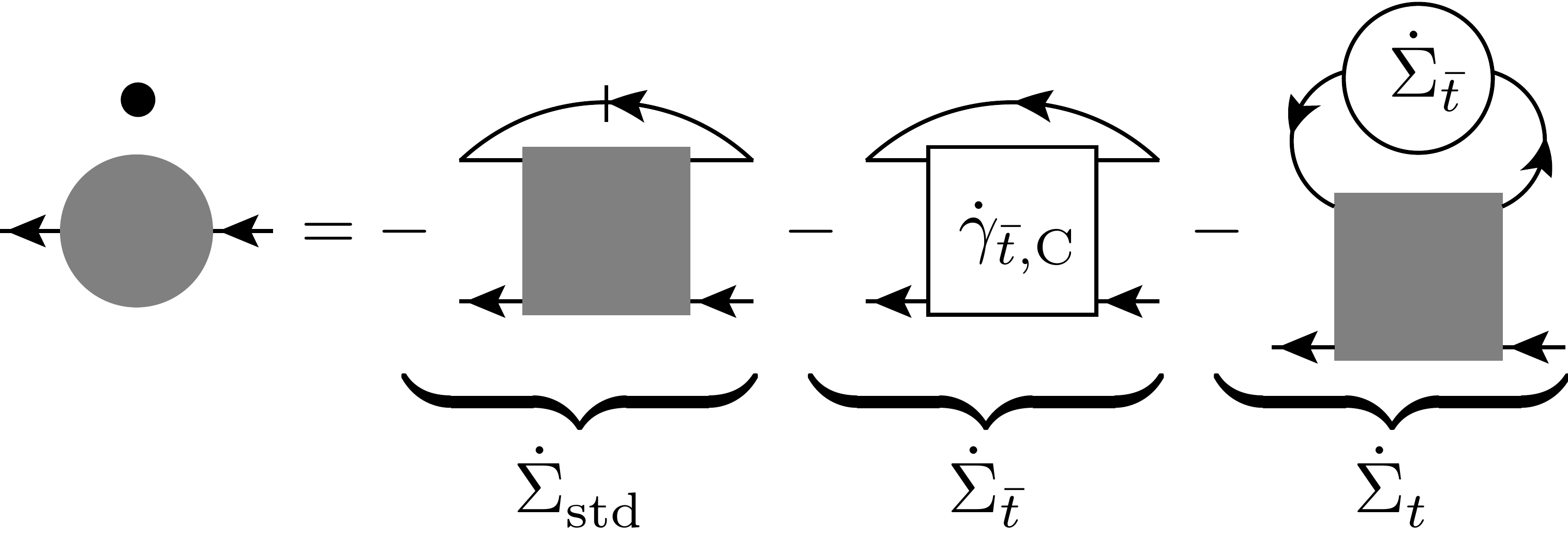}
\caption{%
Multiloop flow equation for the self-energy,
adding two corrections ($\dot{\Sigma}_{\bar{t}}$,
$\dot{\Sigma}_{t}$) to the standard
fRG flow, $\dot{\Sigma}_{\textrm{std}}$.
The (black, thick) line with a dash denotes the
single-scale propagator $S$.%
}
\label{fig:sigma_flow}
\end{figure}
To motivate the addition of $\dot{\Sigma}_{\bar{t}}$ and
$\dot{\Sigma}_{t}$, let us consider the first examples
where multiloop corrections are needed to
complete the derivative of diagrams,
which occur at fourth and fifth order, respectively.
The diagram in \FR{fig:examples}(b)
is obtained by inserting the $\gamma_a$ diagram 
from \FR{fig:examples}(a) (and the symmetry-related $\gamma_t$ diagram)
into the SDE [\FR{fig:sd}(b)].
The differentiated diagram in \FR{fig:examples}(e)
is part of the derivative of \FR{fig:examples}(b),
but not contained in the standard flow.
In fact, the vertex
needed for this diagram to be part of
$\dot{\Sigma}_{\textrm{std}}$
[i.e., the vertex obtained by cutting the differentiated line
in \FR{fig:examples}(e)]
is a so-called envelope vertex,
the lowest-order realization of a nonparquet vertex [cf.~\FR{fig:sd}(b)] 
\footnote{The third-order diagram of $R$ in Fig.~1(b) of Ref.~\onlinecite{Kugler2017} is of nonparquet type only in the X-ray-edge
singularity, where reducibility is required in interband two-particle lines.
The corresponding diagram with identical lines belongs to the $t$ channel.}.
The diagram from \FR{fig:examples}(e) is, however, 
included in the first correction
$\dot{\Sigma}_{\bar{t}}$, as can be seen by inserting
the lowest-order contributions of all vertices 
in the center part
of $\dot{\gamma}_{a}^{(3)}$ 
(using again $\dot{\gamma}_{p}^{(1)}$)
in \FR{fig:vertex_flow}(c) and connecting the top lines.
\begin{figure*}[t]
\center
\includegraphics[width=.99\textwidth]{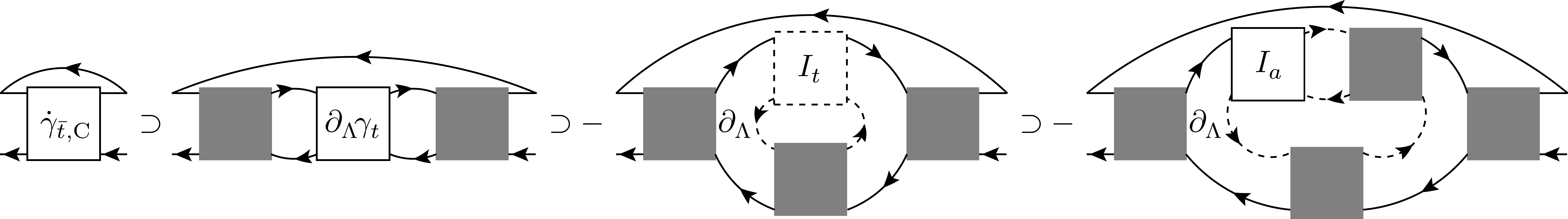}
\caption{%
Special diagrams contributing to $\dot{\Sigma}_{\bar{t}}$.
In the last two diagrams,
we consider a scenario where the differentiated line
is contained in one of the dashed contributions.%
}
\label{fig:sigma_proof1}
\end{figure*}
Inserting the self-energy diagram from \FR{fig:examples}(b)
into the full propagator of the first summand in the SDE [\FR{fig:sd}(b)]
yields the diagram in \FR{fig:examples}(c).
Similar to the previous discussion, one finds that 
the differentiated diagram in \FR{fig:examples}(f),
needed for the full derivative of \FR{fig:examples}(c),
is neither contained in $\dot{\Sigma}_{\textrm{std}}$ 
nor $\dot{\Sigma}_{\bar{t}}$.
It is, however, included in the second mfRG correction, $\dot{\Sigma}_{t}$,
as one of the lowest-order realizations of 
the last summand in \FR{fig:sigma_flow}.
The two extra terms of the mfRG self-energy flow,
$\dot{\Sigma}_{\bar{t}}$ and $\dot{\Sigma}_{t}$,
incorporate the whole multiloop hierarchy of differentiated vertex diagrams via
$\dot{\gamma}_{\bar{t},\textrm{C}}$ [\ER{eq:gamma_tbar_c}]. As is discussed in the 
following subsections, they suffice to generate all parquet diagrams
of $\Sigma$ and, therefore, provide the full dressing of the parquet vertex
in return.
\subsection{Justification}
We will now justify our claim that the mfRG flow
fully generates all parquet diagrams for $\Gamma$ 
and $\Sigma$. We will first show that
all differentiated diagrams in mfRG 
are of the parquet type 
and that there is no overcounting of diagrams.
Concerning the vertex, this has already been done
for the two-channel case of the X-ray-edge singularity \cite{Kugler2017}.
The arguments for the general case are in fact completely
analogous and repeated here for the sake of completeness.
The self-energy is discussed thereafter.
The only totally irreducible contribution to the four-point vertex
in the mfRG flow is the 
bare interaction stemming from the initial condition of the vertex,
$\Gamma_{\Lambda_i}=\Gamma^0$.
All further diagrams on the r.h.s.\ of the flow equations
are obtained by iteratively 
combining two vertices 
by one of the three bubbles from 
\ER{eq:bubbles}.
Hence, they correspond to differentiated \textit{parquet} diagrams
in the respective channel.
The fact that there is no overcounting in mfRG,
i.e., that each diagram occurs at most once,
can be seen employing arguments of diagrammatic reducibility 
and the unique position of the differentiated line in the diagrams.
To be specific, let us consider here the $a$ channel; 
the arguments for the other channels are completely analogous.
First, we note that diagrams in the one-loop term always differ from higher-loop ones.
The reason is that, in higher-loop terms, 
the differentiated line appears in the vertex coming from 
$\partial_{\Lambda} \gamma_{\bar{a}}$. 
This can never contain two vertices connected by an $a$
$G$-$\partial_{\Lambda}G$ bubble, since such terms 
only originate upon differentiating $\gamma_{a}$, the vertex reducible
in $a$ lines.
Second, diagrams in the left, center, or right part
[first, second, and third summand in \FR{fig:vertex_flow}(c), respectively]
of an $\ell$-loop contribution always differ.
This is because the vertex $\gamma^{(\ell)}_{\bar{a}}$
is irreducible in $a$ lines.
The left part is then reducible in $a$ lines \textit{only after} 
the differentiated line appeared,
the right part \textit{only before},
and the center part is reducible in this channel
\textit{before and after} $\partial_{\Lambda}G$.
Third, the same parts (say, the left parts) of different-order
loop contributions ($\ell \neq \ell'$) are always different.
Assume they agreed:
As the $a$ bubble induces the first reducibility 
in this channel, already
$\gamma^{(\ell)}_{\bar{a}}$ and $\gamma^{(\ell')}_{\bar{a}}$ 
would have to agree.
For these, only the same parts can agree, as mentioned before.
The argument then proceeds iteratively until one compares 
the one-loop part to a higher-loop ($|\ell - \ell'| + 1$) one.
These are, however, distinct according to the first point.
Concerning the self-energy,
all diagrams of the flow belong to the parquet type,
since they are constructed from (differentiated)
parquet vertices by closing loops of external legs
in an iterative one-loop procedure.
By cutting one $G^0$ or the $\partial_{\Lambda}G^0$
line in such a self-energy diagram, one can always obtain
a (differentiated) parquet vertex
with possibly dressed amputated legs.
First, there is no overcounting between
$\dot{\Sigma}_{\textrm{std}}$ and
$\dot{\Sigma}_{\bar{t}}$
because cutting the differentiated line
in $\dot{\Sigma}_{\textrm{std}}$ generates a parquet vertex 
(with possibly dressed amputated legs
coming from the single-scale propagator; cf.~\FR{fig:sigma_flow}),
whereas this is not the case for $\dot{\Sigma}_{\bar{t}}$.
To illustrate this statement, we
consider in \FR{fig:sigma_proof1}
a typical case of a $\dot{\Sigma}_{\bar{t}}$ correction,
where we take the $a$ part of $\dot{\gamma}_{\bar{t},\textrm{C}}$
[cf.~\ER{eq:gamma_tbar_c}]
with $\partial_{\Lambda} \gamma_{t}$ in the center. We can insert 
the BSE $\gamma_{t} = B_t(I_t, \Gamma)$ (\FR{fig:bs}) 
and consider simultaneously all scenarios where
the differentiated line, originating from $\partial_{\Lambda} \gamma_{t}$,
is contained in any of the dashed parts.
To be even more specific, we take a specific part
of $I_{t}=R+\gamma_a+\gamma_p$, namely $\gamma_{a}=B_a(I_a,\Gamma)$ (\FR{fig:bs}), 
and consider the cases where the differentiated line, if contained in $I_{t}$, is contained
in the corresponding bubble.
If one now cuts any of the dashed lines,
as candidates for the differentiated line,
one finds that the remaining vertex is \textit{not} of the parquet type,
as it is not reducible in any of the two-particle channels.
The same irreducibility in three lines, when starting to cut the
differentiated line in $\dot{\gamma}_{\bar{t},\textrm{C}}$, occurs in all 
diagrammatic realizations of $\dot{\Sigma}_{\bar{t}}$.
Since the standard flow $\dot{\Sigma}_{\textrm{std}}$
with the \textit{full} instead of the \textit{parquet} vertex is exact,
it follows that the $\dot{\Sigma}_{\bar{t}}$ part
can be written similarly as $\dot{\Sigma}_{\textrm{std}}$,
but using a \textit{nonparquet} (np) vertex [\FR{fig:sigma_proof2}(a)].
As a consequence, $\dot{\Sigma}_{t}$, obtained by connecting 
$\dot{\Sigma}_{\bar{t}}$ and $\Gamma$ by a $t$ bubble,
can similarly be written with a nonparquet vertex [\FR{fig:sigma_proof2}(b)].
Thus, there cannot be any overcounting between 
$\dot{\Sigma}_{\textrm{std}}$ and
$\dot{\Sigma}_{t}$, either.
Finally, there is likewise no overcounting between
$\dot{\Sigma}_{\bar{t}}$ and $\dot{\Sigma}_{t}$:
After removing the differentiated line in 
$\dot{\Sigma}_{\bar{t}}$, the remaining nonparquet vertex 
$\Gamma_{\textrm{np}}$  is in particular irreducible in the $t$ channel
(as was discussed above). However, removing the differentiated line in 
$\dot{\Sigma}_{t}$ after expressing 
$\dot{\Sigma}_{\bar{t}}$ via $\Gamma_{\textrm{np}}$ [cf.~\FR{fig:sigma_proof2}(b)],
the remaining vertex $\Gamma'_{\textrm{np}}$ 
is by construction reducible 
in $t$ lines (although not a parquet vertex).
\begin{figure}[t]
\center
\includegraphics[width=.48\textwidth]{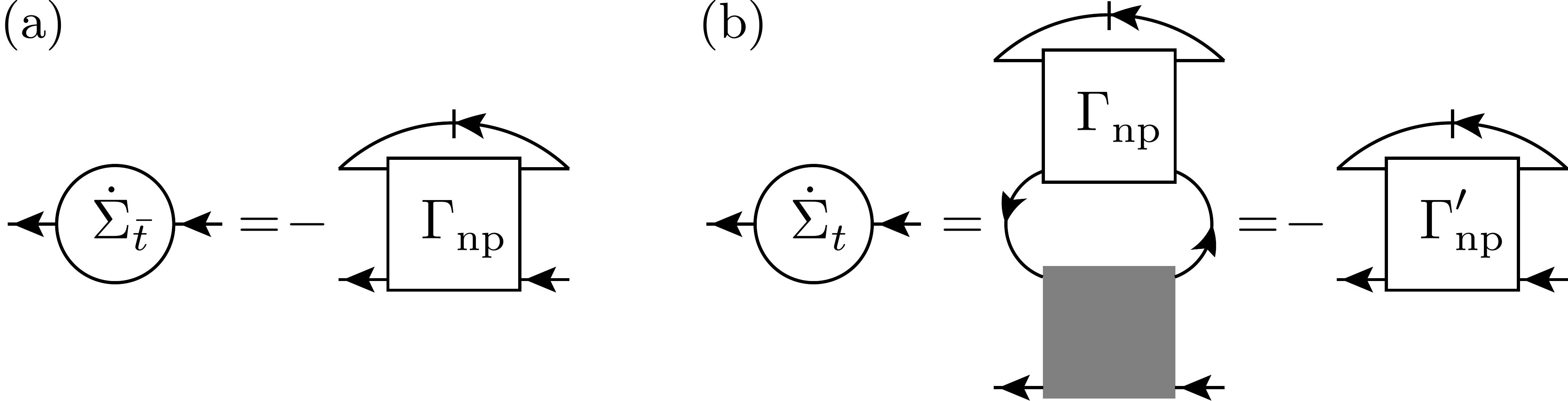}
\caption{%
Rewriting of the corrections to the self-energy flow:
(a) $\dot{\Sigma}_{\bar{t}}$ can be expressed by
a nonparquet vertex $\Gamma_{\textrm{np}}$
contracted with the single-scale propagator $S$.
(b) $\dot{\Sigma}_{t}$, obtained by connecting 
$\dot{\Sigma}_{\bar{t}}$ and $\Gamma$ by a $t$ bubble,
then involves a bubble connecting a nonparquet and parquet vertex,
which yields another nonparquet vertex $\Gamma'_{\textrm{np}}$,
contracted with $S$.%
}
\label{fig:sigma_proof2}
\end{figure}
In summary, all diagrams
of the four-point vertex and self-energy
generated by the mfRG flow
belong to the parquet class
and are included at most once. 
To show that the mfRG flow generates \textit{all} differentiated parquet diagrams,
we will demonstrate next that, at any given order in the interaction,
their number is equal to the number of diagrams generated by the mfRG flow.
\subsection{Counting of diagrams}
\label{sec:diagr_count}
In order to count the number of diagrams in all involved functions,
we make use of either exact, self-consistent equations
or the mfRG flow equations.
As a first example, we count the number of diagrams 
in the full propagator $G$ at order $n$ in the interaction, $\N_G(n)$,
given the number of diagrams in the self-energy, $\N_{\Sigma}(n)$.
Concerning the bare propagator and self-energy,
we know
$\N_{G^0}(n) = \delta_{n,0}$ and
$\N_{\Sigma}(0) = 0$.
From Dyson's equation \ERn{eq:dyson}, we then get
\begin{equation}
\N_G(n) = \delta_{n,0} + \sum_{m=1}^{n} \N_{\Sigma}(m) \N_G(n-m)
\ED
\label{eq:NG}
\end{equation}
Defining a convolution of sequences, according to
\begin{equation}
\N_1 = \N_2 \ast \N_3
\ \Leftrightarrow \
\N_1(n) = \sum_{m=0}^{n} \N_2(m) \N_3(n-m)
\ \forall n
\EC
\end{equation}
we can write \ER{eq:NG} in direct analogy to the 
original equation \ERn{eq:dyson} as
\begin{equation}
\N_G = \N_{G^0} + \N_{G^0} \ast \N_{\Sigma\pss} \ast \N_{G\pss}
\ED
\end{equation}
Similar relations for the self-energy and vertex 
can be obtained from the SDE \ERn{eq:SchwingerDyson},
the parquet equation \ERn{eq:parquet}, and 
the BSEs \ERn{eq:BetheSalpeter}.
The number of diagrams in the bare vertex
is $\N_{\Gamma^0}=\delta_{n,1}$
(one can also take any $\N_{\Gamma^0} \propto \delta_{n,1}$).
From the SDE \ERn{eq:SchwingerDyson}, we get for the self-energy
\begin{align}
\N_{\Sigma} = \N_{\Gamma^0} \ast \N_{G\pss}
+ \tfrac{1}{2}\,
\N_{\Gamma^0} \ast \N_{G\pss} \ast \N_{G\pss} \ast \N_{G\pss} \ast \N_{\Gamma\pss}
\ED
\label{eq:diagr_count_sd}
\end{align}
Note that, when counting diagrams, we can ignore the extra minus signs
but must keep track of prefactors of magnitude not equal to unity.
These prefactors avoid double counting 
of the antisymmetric vertex
\cite{Bickers2004}
and originate from the way the diagrams are constructed
\footnote{In the SDE \ERn{eq:SchwingerDyson}, e.g., the self-energy is constructed in a way
that involves two parallel lines connected to the same vertex,
requiring a factor of $1/2$ to avoid double counting. 
In the standard flow equation \ERn{eq:sigma_flow_std} for $\Sigma$, 
no such lines exist and hence no extra factor either.}.
Concerning the full vertex,
we can use 
that the symmetry relation between 
the $a$ and $t$ bubble given 
in \ER{eq:a_t_sym} holds for the full
reducible vertices $\gamma_{a}$ and $\gamma_{t}$ \cite{Bickers2004},
such that $\N_{\gamma_{a}}=\N_{\gamma_{t}}$.
In the parquet approximation $R=\Gamma^0$, and
the parquet equation \ERn{eq:parquet} and the BSEs \ERn{eq:BetheSalpeter} yield
\begin{subequations}
\begin{align}
\N_{\Gamma} 
& = 
\N_{R} + 2\, \N_{\gamma_a} + \N_{\gamma_p}
\label{eq:diagr_count_R}
\\
\N_{\gamma_a} 
& = 
( \N_{\Gamma} - \N_{\gamma_a} ) \ast \N_G \ast \N_G \ast \N_{\Gamma}
\label{eq:diagr_count_gamma_a} \\
\N_{\gamma_p} 
& = 
\tfrac{1}{2} ( \N_{\Gamma} - \N_{\gamma_p} ) \ast \N_G \ast \N_G \ast \N_{\Gamma}
\ED
\label{eq:diagr_count_gamma_p}
\end{align}
\end{subequations}
Since $\N_{\Gamma^0}(0) = 0$, these equations,
just like the original equations, can be solved iteratively.
Knowing the number of diagrams in 
all quantities up to order $n-1$
allows one to calculate them at order $n$.
This can also be done numerically. 
Table \ref{tab:num_diagr} (first two lines) shows the number of
parquet diagrams up to order 6.
For large interaction order $n$, we find that
the number of diagrams in the parquet vertex
and self-energy grows exponentially
in $n$ [cf.\ \FR{fig:diagrcount}(a)].
To prove our claim that the mfRG flow generates
all parquet diagrams, we must count the number of
diagrams, $\N_{\dot{\Sigma}}(n)$ and $\N_{\dot{\gamma}_r}(n)$,
obtained by differentiating the set of all 
corresponding parquet graphs.
Then, we check that these numbers are exactly reproduced 
by the number of diagrams contained on the r.h.s.\
of the mfRG flow equations.
A diagram of the full propagator 
at order $n$ has $2n+1$ internal lines,
a self-energy diagram $2n-1$,
and vertex diagram $2n-2$.
According to the product rule, the number of differentiated
diagrams is thus
\begin{subequations}
\begin{align}
\N_{\dot{G}}(n) & = \N_G(n) (2n+1)
\EC \\
\N_{\dot{\Sigma}}(n) & = \N_{\Sigma}(n) (2n-1)
\label{eq:diagr_count_sigma_dot}
\EC \\
\N_{\dot{\gamma}_r}(n) & = \N_{\gamma_r}(n) (2n-2)
\ED
\end{align}
\end{subequations}
\begin{table}[t]
\begin{tabular*}{0.48\textwidth}{@{\extracolsep{\fill}} l  c  c  c  c  c  c }
\hline\hline
\rule{0pt}{3ex}
$n$ & 1 & 2 & 3 & 4 & 5 & 6 
\\
\\[-3.0ex]
\hline
\rule{0pt}{3ex}%
$\N_{\Gamma}$ & 1 & 2$\frac{1}{2}$ & 15$\tfrac{1}{4}$ & 108$\tfrac{1}{8}$ & 832$\tfrac{1}{16}$ & 6753$\tfrac{21}{32}$
\\ 
$\N_{\Sigma}$ & 1 & 1$\frac{1}{2}$ & 5$\frac{1}{4}$ & 25$\frac{7}{8}$ & 156$\frac{1}{16}$ & 1073$\frac{3}{32}$
\\ 
\rule{0pt}{4ex}%
$\N_{\dot{\Gamma}}$ & 0 & 5 & 61 & 648$\frac{3}{4}$ & 6656$\frac{1}{2}$ & 67536$\frac{9}{16}$
\\ 
$\N_{\dot{\Gamma}^{(1\ell)}}$ & 0 & 5 & 45 & 373$\frac{3}{4}$ & 3117$\frac{1}{2}$ & 26519$\frac{1}{16}$
\\ 
$\N_{\dot{\Gamma}^{(2\ell)}}$ & 0 & 0 & 16 & 216 & 2264 & 21972
\\ 
$\N_{\dot{\Gamma}^{(3\ell)}}$ & 0 & 0 & 0 & 59 & 1062 & 13481$\frac{1}{2}$
\\ 
$\N_{\dot{\Gamma}^{(4\ell)}}$ & 0 & 0 & 0 & 0 & 213 & 4792$\frac{1}{2}$
\\ 
$\N_{\dot{\Gamma}^{(5\ell)}}$ & 0 & 0 & 0 & 0 & 0 & 771$\frac{1}{2}$
\\ 
\rule{0pt}{4ex}%
$\N_{\dot{\Sigma}}$ & 1 & 4$\frac{1}{2}$ & 26$\frac{1}{4}$ & 181$\frac{1}{8}$ & 1404$\frac{9}{16}$ & 11804$\frac{1}{32}$
\\ 
$\N_{\dot{\Sigma}_{\textrm{std}}}$ & 1 & 4$\frac{1}{2}$ & 26$\frac{1}{4}$ & 177$\frac{1}{8}$ & 1311$\frac{9}{16}$ & 10348$\frac{1}{32}$
\\ 
$\N_{\dot{\Sigma}_{\bar{t}}}$ & 0 & 0 & 0 & 4 & 89 & 1349
\\ 
$\N_{\dot{\Sigma}_{t}}$ & 0 & 0 & 0 & 0 & 4 & 107
\\
\\[-3.0ex]
\hline\hline
\end{tabular*}
\caption{Number of (bare) parquet diagrams, differentiated parquet diagrams, and diagrams generated by mfRG up to interaction order 6 and loop order 5.
Fractional parts originate from multiple factors of $1/2$, used to
avoid double counting of the antisymmetric vertex \cite{Bickers2004}.
As we use $\N_{\Gamma^0}=\delta_{n,1}$, we count
Hugenholtz diagrams 
\cite{Negele2008} [where, e.g., $\N_{\Sigma}(1)=1$, cf.\ \FR{fig:dyson}].
The choice $\N_{\Gamma^0}=2\delta_{n,1}$ [cf.\ \ER{eq:vertex_U}] would give an extra factor $2^n$
for all numbers of diagrams at order $n$, resulting in the (integer) numbers 
of Feynman diagrams [where, e.g., $\N_{\Sigma}(1)=2$].%
}
\label{tab:num_diagr}
\end{table}
From the mfRG flow of the vertex [\ER{eq:multiloop_flow}], 
we deduce
\begin{subequations}
\begin{align}
\N_{\dot{\gamma}_{a}^{(1)}} 
& = 2\, \N_{\Gamma\pd} \ast \N_{\dot{G}} \ast \N_{G\pd} \ast \N_{\Gamma\pd}
\EC \\
\N_{\dot{\gamma}_{p}^{(1)}} 
& = \N_{\Gamma\pd} \ast \N_{\dot{G}} \ast \N_{G\pd} \ast \N_{\Gamma\pd}
\EC \\
\N_{\dot{\gamma}_{a}^{(2)}} 
& = 2\, ( \N_{\dot{\gamma}^{(1)}_a} + \N_{\dot{\gamma}^{(1)}_p} ) 
\ast \N_{\Pi} \ast \N_{\Gamma}
\EC \\
\N_{\dot{\gamma}_{p}^{(2)}} 
& = 2\, \N_{\dot{\gamma}^{(1)}_a} \ast \N_{\Pi} \ast \N_{\Gamma}
\EC
\end{align}
\end{subequations}
where $\N_{\Pi} = \N_G \ast \N_G$
denotes the number of diagrams in a bubble.
For $\ell+2 \geq 3$, we have
\begin{subequations}
\begin{align}
\N_{\dot{\gamma}_{a}^{(\ell+2)}} 
& = 2\, ( \N_{\dot{\gamma}^{(\ell+1)}_a} + \N_{\dot{\gamma}^{(\ell+1)}_p} ) 
\ast \N_{\Pi} \ast \N_{\Gamma}
\nonumber  \\
& \ +
\N_{\Gamma} \ast \N_{\Pi} \ast
( \N_{\dot{\gamma}^{(\ell)}_a} + \N_{\dot{\gamma}^{(\ell)}_p} ) 
\ast \N_{\Pi} \ast \N_{\Gamma}
\EC \\
\N_{\dot{\gamma}_{p}^{(\ell+2)}} 
& = 2\, \N_{\dot{\gamma}^{(\ell+1)}_a} 
\ast \N_{\Pi} \ast \N_{\Gamma}
\nonumber \\
& \ +
\tfrac{1}{2} \,
\N_{\Gamma} \ast \N_{\Pi} \ast
\N_{\dot{\gamma}^{(\ell)}_a}
\ast \N_{\Pi} \ast \N_{\Gamma}
\ED
\end{align}
\end{subequations}
Summing all loop contributions yields
\begin{equation}
\textstyle
\N_{\dot{\gamma}_a}^{\textrm{mfRG}}
= 
\sum_{\ell \geq 1} \N_{\dot{\gamma}^{(\ell)}_a} 
\EC
\quad
\N_{\dot{\gamma}_p}^{\textrm{mfRG}}
= 
\sum_{\ell \geq 1} \N_{\dot{\gamma}^{(\ell)}_p} 
\ED
\end{equation}
For the flow of the self-energy 
\ERn{eq:sigma_flow}, we need the
center part of the vertex flow in the $a$ and $p$ channel,
for which the number of diagrams sums up to
\begin{align}
\N_{\dot{\gamma}_{\bar{t},\textrm{C}}} 
& = 
\N_{\Gamma} \ast \N_{\Pi} \ast
\Big( \tfrac{3}{2} \, 
\N_{\dot{\gamma}_a}^{\textrm{mfRG}} + \N_{\dot{\gamma}_p}^{\textrm{mfRG}}
\Big) 
\ast \N_{\Pi} \ast \N_{\Gamma}
\ED
\end{align}
The number of diagrams in the single-scale propagator $S$
\ERn{eq:singlescale}
can be obtained from two equivalent relations
\begin{subequations}
\begin{align}
\N_{S\pd} 
& = 
\N_{\dot{G}} - \N_{G\pd} \ast N_{\dot{\Sigma}} \ast \N_{G\pd}
\\
& = 
( \N_{\mathbbm{1}\pd} + \N_{G\pd} \ast \N_{\Sigma\pd} ) \ast \N_{\dot{G}^0} \ast
( \N_{\mathbbm{1}\pd} + \N_{\Sigma\pd} \ast \N_{G\pd} )
\EC
\end{align}
\end{subequations}
with
$\N_{\dot{G}^0}(n) = \delta_{n,0} = \N_{\mathbbm{1}}(n)$.
From \ER{eq:sigma_flow}, we then get
\begin{flalign}
\N_{\dot{\Sigma}}^{\textrm{mfRG}}
& = 
\N_{\dot{\Sigma}_{\textrm{std}}} + \N_{\dot{\Sigma}_{\bar{t}}} 
+ \N_{\dot{\Sigma}_{t}}
\EC
&
\N_{\dot{\Sigma}_{\textrm{std}}} 
& = 
\N_{\Gamma\pd} \ast \N_{S\pd}
\EC
\nonumber \\
\N_{\dot{\Sigma}_{\bar{t}}} 
& = 
\N_{\dot{\gamma}_{\bar{t},\textrm{C}}}  \ast \N_{G\pd}
\EC
&
\N_{\dot{\Sigma}_{t}}
& = 
\N_{\Gamma\pd} \ast \N_{\Pi\pd} \ast \N_{\dot{\Sigma}_{\bar{t}}}
\ED
\label{eq:diagr_count_sigma_flow}
\end{flalign}
\begin{figure}[t]
\includegraphics[width=.48\textwidth]{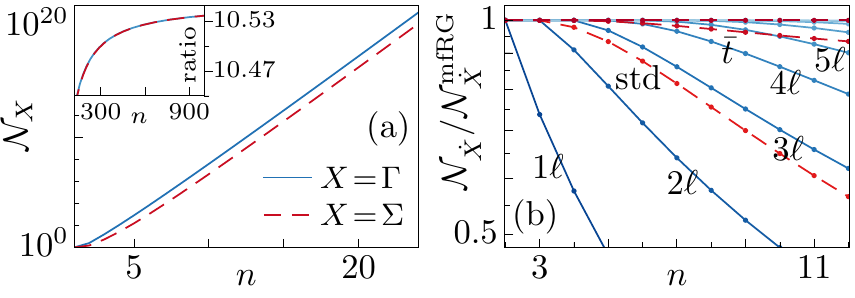}
\caption{%
Logarithmic plots for the number of diagrams at
interaction order $n$ for both
vertex and self-energy.
(a)
$\N_{\Gamma}$, $\N_{\Sigma}$ 
grow exponentially for large $n$ (inset:
the ratio of subsequent elements approaches a constant).
(b)
The cumulative low-loop vertex flows 
($1\ell$ up to $5\ell$)
and the 
self-energy flows $\dot{\Sigma}_{\textrm{std}}$ (labeled std) and 
$\dot{\Sigma}_{\textrm{std}}+\dot{\Sigma}_{\bar{t}}$ (labeled $\bar{t}$) 
miss differentiated parquet diagrams.
However, the full multiloop flow for
vertex and self-energy
generates all differentiated parquet diagrams 
to arbitrary order in the interaction.%
}
\label{fig:diagrcount}
\end{figure}
Numerically, one can check order for order in the interaction
[cf.~Table \ref{tab:num_diagr} and \FR{fig:diagrcount}(b)]
that, indeed, the mfRG flow generates exactly the same
number of diagrams as obtained by differentiating 
all parquet diagrams, i.e.,
\begin{align}
\N_{\dot{\gamma}_r}^{\textrm{\vphantom{G}}}(n)
& = 
\N_{\dot{\gamma}_r}^{\textrm{mfRG}}(n)
\EC \quad
\N_{\dot{\Sigma}}^{\textrm{\vphantom{G}}}(n) 
= 
\N_{\dot{\Sigma}}^{\textrm{mfRG}}(n) 
\quad
\forall n
\ED
\end{align}
This demonstrates the equivalence between solving
the multiloop fRG flow and solving the (first-order)
parquet equations for a general model.
\subsection{Computational aspects}
\label{sec:mfRG_comp}
All contributions to the mfRG flow---%
for the vertex as well as for the self-energy---%
are of an iterative one-loop structure
and hence well suited for numerical algorithms.
In fact, by keeping track of the left (L) and right (R) summands
in the higher-loop vertex flow \ERn{eq:higher-loop_flow}
\begin{equation}
\dot{\gamma}_{r,\textrm{L}}^{(\ell+2)}
= 
B_r \big( \gamma_{\bar{r}}^{(\ell+1)}, \Gamma \big)
\EC \quad
\dot{\gamma}_{r,\textrm{R}}^{(\ell+2)}
= 
B_r \big( \Gamma, \gamma_{\bar{r}}^{(\ell+1)} \big)
\EC
\end{equation}
the center part \ERn{eq:vertex_center_part} can be efficiently computed as
\begin{equation} 
\dot{\gamma}_{r,\textrm{C}}^{(\ell+2)}
= 
B_r \big( \Gamma, \gamma_{r,\textrm{L}}^{(\ell+1)} \big)
=
B \big( \gamma_{r,\textrm{R}}^{(\ell+1)}, \Gamma \big)
\ED
\end{equation}
Consequently, the numerical effort in the multiloop corrections
of the vertex flow scales linearly in $\ell$.
The self-energy flow \ERn{eq:sigma_flow}
is already stated with one integration only.

The (standard) fRG hierarchy of flow equations constitutes
a (first-order) ordinary differential equation.
Neglecting the six-point vertex, it can be written as
\begin{equation}
\partial_{\Lambda} \Sigma = f_{\Sigma}^{\textrm{std}} (\Lambda, \Sigma, \Gamma)
\EC \quad
\partial_{\Lambda} \Gamma = f_{\Gamma}^{\textrm{std}} (\Lambda, \Sigma, \Gamma)
\EC
\end{equation}
where, here and henceforth, $f$ denotes the part of the r.h.s.\ of 
the flow equation corresponding to its indices.
Improving this approximation by
adding differentiated self-energy contributions in the
vertex flow (as is also done in  mfRG), $f_{\Gamma}^{\textrm{std}}$
is replaced by another function
$\tilde{f}_{\Gamma}^{\textrm{std}} (\Lambda, \Sigma, \Gamma, \partial_{\Lambda} \Sigma)$,
which further depends on the $\Lambda$ derivative of the self-energy.
Such a differential equation is still feasible
for many algorithms as one can simply compute 
$\partial_{\Lambda} \Sigma$ first and use it in the
calculation of $\partial_{\Lambda} \Gamma$.
However, the full mfRG flow for the vertex and self-energy
has the form
\begin{equation}
\partial_{\Lambda} \Sigma = 
f_{\Sigma} (\Lambda, \Sigma, \Gamma, \partial_{\Lambda} \Gamma)
\EC \quad
\partial_{\Lambda} \Gamma = 
f_{\Gamma} (\Lambda, \Sigma, \Gamma, \partial_{\Lambda} \Sigma)
\EC
\end{equation}
in which derivatives occur on all parts of the r.h.s.,
yielding an algebraic (as opposed to ordinary)
differential equation.
Techniques to solve algebraic differential equations 
exist, but a discussion of them exceeds the scope of this paper.
Let us merely suggest an approximate solution strategy
that reduces the mfRG flow to an ordinary differential equation,
has no computational overhead, and deviates from the exact flow
starting at sixth order in the interaction,
summarized as follows:
\begin{subequations}
\label{eq:approx_mfRG_sol}
\begin{align}
\dot{\Sigma}_{\textrm{std}} 
& = 
f_{\dot{\Sigma}_{\textrm{std}}} (\Lambda, \Sigma, \Gamma)
\EC
\label{eq:approx_mfRG_sol_1} \\
\partial_{\Lambda} \Gamma
& \approx
\dot{\Gamma}_{\textrm{approx}}
=
f_{\Gamma} 
(\Lambda, \Sigma, \Gamma, \partial_{\Lambda} \Sigma=\dot{\Sigma}_{\textrm{std}} )
\EC
\label{eq:approx_mfRG_sol_2} \\
\partial_{\Lambda} \Sigma
& \approx
\dot{\Sigma}_{\textrm{std}} +
f_{\dot{\Sigma}_{\bar{t}}}
(\Lambda, \Sigma, \partial_{\Lambda} \Gamma = \dot{\Gamma}_{\textrm{approx}})
\nonumber \\ & \ +
f_{\dot{\Sigma}_{t}}
(\Lambda, \Sigma, \partial_{\Lambda} \Gamma = \dot{\Gamma}_{\textrm{approx}})
\ED
\label{eq:approx_mfRG_sol_3}
\end{align}
\end{subequations}
According to this scheme, one computes first the standard flow of the self-energy,
which deviates from the full $\Sigma$ flow at interaction order $U^4$.
Inserting this into the vertex flow yields an approximate
vertex derivative, $\dot{\Gamma}_{\textrm{approx}}$,
where deviations from the full flow,
induced by the approximate form of $\partial_{\Lambda} \Sigma$,
start at order $U^6$.
The center part of the vertex flow involves at least four vertices, 
such that deviations, induced by the self-energy, start at order $U^8$.
The resulting, approximate $\dot{\gamma}_{\bar{t},\textrm{C}}$ 
can then be used to complete $\partial_{\Lambda} \Sigma$, adding the terms 
$\dot{\Sigma}_{\bar{t}}$ and $\dot{\Sigma}_{t}$,
such that the self-energy flow is correctly computed
up to errors of order $U^8$.
Evidently, this scheme can also be iterated
[using \EsR{eq:approx_mfRG_sol_2} and \ERn{eq:approx_mfRG_sol_3}], 
increasing the accuracy
by four orders with each step.
We have attached a pseudocode for such a solution strategy of the mfRG flow 
in \AR{appendix1}.
\subsection{{General aspects}}
Since the standard fRG flow for the
self-energy and four-point vertex---\textit{including} the
six-point vertex---is exact,
all mfRG corrections can be understood
as fully simulating the effect of the six-point
vertex on parquet diagrams of $\Sigma$ and $\Gamma$.
For instance, the two-loop corrections to the
vertex flow and the Katanin substitution in the 
improved one-loop flow equation contain all
third-order contributions of the six-point vertex 
\cite{Kugler2017a, Katanin2004, Eberlein2014}.
Nevertheless, in the standard fRG hierarchy of flow equations,
the parquet graphs comprise $n$-point vertices of 
arbitrary order ($n$) \cite{Kugler2017a},
such that a non-diagrammatic derivation of mfRG based on this hierarchy appears rather difficult.
Conversely, the derivation of the mfRG flow does not rely on the
fRG hierarchy or properties of the (quantum) effective action;
it can thus be understood independently and without prior knowledge
of fRG.
The mfRG flow at the two- or higher-loop level
is exact up to third
order in the interaction and therefore naturally fulfills Ward identities
with accuracy $\textit{O}(\Gamma^4)$, compared to
$\textit{O}(\Gamma^3)$ in the case of one-loop fRG \cite{Katanin2004}.
Yet, since the parquet self-energy is exact up to \textit{fourth} order
but the parquet vertex only up to \textit{third} order, such identities
are typically violated starting at fourth order.
One can think of schemes to extend mfRG beyond the parquet 
approximation. However, we find those rather impracticable
and only briefly mention them in \AR{appendix2}.
Furthermore, the mfRG flow is applicable for any initial condition of the vertex functions.
Whereas the choice $G_{\Lambda_i}=0$ used here leads to a summation of all parquet diagrams,
starting the mfRG flow from the local quantities of 
dynamical mean-field theory (DMFT) \cite{Georges1996,Taranto2014}
allows one to add nonlocal correlations, similarly to
solving the parquet equations in the dynamical vertex approximation 
(D$\Gamma$A) \cite{Toschi2007, Held2008, Valli2010}.
However, contrary to D$\Gamma$A, the mfRG flow is built on the full vertex $\Gamma^{(4)}_{\textrm{DMFT}}$
and does not require the \textit{diagrammatic}
decomposition of the \textit{nonperturbative} vertex
\footnote{Alternatives to D$\Gamma$A which do not require the totally irreducible vertex are the dual fermion \cite{Rubtsov2008,Brener2008,Hafermann2009} and the related 1PI approach \cite{Rohringer2013}. However, upon transformation to the dual variables, the bare action contains $n$-particle vertices for all $n$. Recent studies \cite{Ribic2017,Ribic2017a} show that the corresponding six-point vertex yields sizable contributions for the (physical) self-energy, and it remains unclear how a truncation in the (dual) bare action can be justified.}
\nocite{Rubtsov2008,Brener2008,Hafermann2009,Rohringer2013,Ribic2017,Ribic2017a}
$\Gamma^{(4)}_{\textrm{DMFT}} = R + \sum_r \gamma_r$
that leads to diverging results close
to a quantum phase transition \cite{Schaefer2013,Schaefer2016,Gunnarsson2017}.
Inspecting the one-loop flow equations of the vertex once more, 
we observe that diagrams on the r.h.s.\
contain the differentiated propagator \textit{only} 
in the two-particle lines that induce the reducibility.
Propagators which appear in two-particle lines
which do not induce the reducibility are not differentiated.
Therefore,
only those diagrams that are reducible in \textit{all} positions of
two-particle lines---the so-called ladder diagrams---are
fully included.
It follows that the standard truncated, one-loop fRG flow is biased towards 
\textit{ladder} constructions of the four-point vertex.
For a constant interaction $U$ and a transfer energy-momentum $\Omega$, ladder diagrams 
of a certain channel can easily be summed to
$\Gamma^{\textrm{ladder}}_{\Omega} = U ( 1 - U \Pi_{\Omega} )^{-1}$,
where $\Pi_{\Omega}$ is the corresponding bubble.
Ladder diagrams are therefore particularly prone to divergences
with increasing $U$ or increasing values of $\Pi_{\Omega}$
(as can occur upon lowering the cutoff scale $\Lambda$)
and can thus be responsible for
premature vertex divergences in fRG.
Indeed, so far, 
fRG computations have often suffered from such vertex divergences,
and the flow had be stopped at finite RG scale $\Lambda_c$
\cite{Metzner2012, Eberlein2014a}.
In this context, the two-loop corrections have 
been found to significantly reduce
the critical scale of vertex divergences $\Lambda_c$ \cite{Eberlein2014,Rueck2017}. 
This suggests that it would be worthwhile to study the effect 
of higher-loop mfRG corrections---we expect that they reduce
$\Lambda_c$ even further.
Throughout this paper, we have taken a perspective
that views fRG as a tool to resum diagrams (say, \textit{physical} diagrams)
by integrating a collection of \textit{differentiated} (and thus
$\Lambda$-dependent) diagrams. In this regard,
the mfRG corrections do not add new \textit{physical} diagrams to the flow,
they only add \textit{differentiated} diagrams 
to complete those derivatives of physical diagrams that are only
partially contained by one-loop fRG.
In other words, for any physical diagram 
to which a differentiated diagram of mfRG contributes,
there also exists a differentiated diagram in one-loop fRG.
The differentiated diagrams of the higher-loop
corrections \textit{and} the one-loop flow 
all contribute the \textit{same} set of 
physical diagrams---the parquet diagrams.
Whereas the \textit{one-loop} flow of the vertex contains differentiated propagators
at the two-particle-reducible positions, the \textit{multiloop} flow iteratively
adds those parts for which the differentiated line is increasingly nested.
Such non-ladder contributions are crucial to suppress vertex divergences
originating from the summation of ladder diagrams \cite{Kugler2017}.
Similarly, the \textit{standard} self-energy flow does not form a total derivative any more
if one has only the parquet vertex at one's disposal.
All diagrams of the standard flow are of the parquet type,
but differentiated lines in heavily nested positions are omitted (cf.~\FR{fig:examples}).
The mfRG corrections incorporate all remaining
contributions by two additions that build up on the multiloop vertex flow.
Altogether, the mfRG flow achieves a full summation
of all parquet diagrams of the vertex and self-energy.
Consequently, mfRG solutions are no longer dependent on
the specific way the $\Lambda$ dependence (regulator) was introduced
\cite{Kugler2017} and thus fully implement the 
meaning of the original fRG idea.
\vspace{-0.1cm}
\section{Conclusion}
\label{sec:conclusion}
We have presented multiloop fRG flow equations for the
four-point vertex and self-energy, formulated for the general 
fermionic many-body problem. 
The mfRG corrections fully simulate the effect 
of the six-point vertex on parquet diagrams,
completing the derivatives of diagrams that are only partially contained in
the standard truncated fRG flow.
Whereas one-loop fRG contains differentiated propagators only
at the two-particle-reducible positions and the standard
self-energy flow does not suffice to form a total derivative when having
only the parquet vertex at one's disposal,
the multiloop iteration adds all remaining parts, 
where the differentiated line appears at increasingly nested positions.
We have motivated the multiloop corrections at low orders
and ruled out any overcounting of diagrams.
Moreover, we have put forward a simple recipe to count diagrams
and numerically check that the mfRG flow generates 
all differentiated parquet diagrams
for the vertex and self-energy, order for order in the interaction.
Due to its iterative one-loop structure, the mfRG flow
is well suited for efficient numerical computations.
We have given a simple approximation, which
renders the algebraic differential equation
accessible to standard solvers for ordinary 
differential equations and exhibits
only minor deviations from
the full mfRG flow.
Given the general formulation,
the benefits of mfRG on physical problems
can be exploited in a large number of fRG applications.
The full resummation of parquet diagrams via mfRG
eliminates the bias of fRG computations towards
divergent ladder constructions of the vertex
and restores the independence on the choice of regulator.
We expect that this will generically enhance the usefulness
of the truncated fRG framework and 
increase the robustness of the physical conclusions
drawn from fRG results.
\vspace{-0.2cm}
\begin{acknowledgments}
We thank S.\ Jakobs for 
pointing out the need for multiloop 
corrections to the self-energy flow
and W.\ Metzner and A.\ Toschi for useful discussions.
We acknowledge support by the Cluster of Excellence
Nanosystems Initiative Munich;
F.B.K.\ acknowledges funding from
the research school IMPRS-QST.
\end{acknowledgments}
\appendix
\section{Pseudocode implementation}
\label{appendix1}
%

%
\begin{alg-table}[b]
\begin{algorithmic}[1]
\Statex \textbf{ Function} $f(\Lambda, \Psi$):

\Indp
\setcounter{ALG@line}{0}
\State $S = S(\Lambda,\Psi.\Sigma)$
\State $G = G(\Lambda,\Psi.\Sigma)$
\State $\dif\Sigma_{\textrm{std}} = L( \Psi.\Gamma, S) $
\State $\dif\Psi.\Sigma = \dif\Sigma_{\textrm{std}} $

\FOR{$it = 1 \dots it_f$}
\State $\dif G = S + G\cdot \dif\Psi.\Sigma \cdot G$
\FOR{$r=a,p,t$}
\State $\dif\gamma_r = \dot{B}_r(\Psi.\Gamma, \Psi.\Gamma, G, \dif G )$
\ENDFOR

\Statex /* \textit{jump to line 41 for one-loop fRG} */

\FOR{$r=a,p,t$}
\State $\dif\gamma_r^{\textrm{L}} = B_r\big( \sum_{r'\neq r}\dif\gamma_{r'}, 
	\Psi.\Gamma, G \big)$
\State $\dif\gamma_r^{\textrm{R}} = B_r\big( \Psi.\Gamma, 
	\sum_{r'\neq r}\dif\gamma_{r'}, G \big)$
\ENDFOR
\FOR{$r=a,p,t$}
\State $\dif\gamma_r^{\textrm{T}} = \dif\gamma_r^{\textrm{L}} + \dif\gamma_r^{\textrm{R}}$
\State $\dif\gamma_r \leftarrow \dif\gamma_r + \dif\gamma_r^{\textrm{T}}$
\ENDFOR

\Statex /* \textit{jump to line 41 for two-loop fRG} */

\State $\dif\gamma_{\bar{t}}^{\textrm{C}} = 0$
\FOR{$\ell=3 \dots \ell_f$}
\FOR{$r=a,p,t$}
\State $\dif\gamma_r^{\textrm{C}} = B_r( \Psi.\Gamma, \dif\gamma_{r}^{\textrm{L}}, 
	G )$
\State $\dif\gamma_r^{\textrm{L}} = B_r\big( \sum_{r'\neq r}\dif\gamma_{r'}^{\textrm{T}}, \Psi.\Gamma, G \big)$
\State $\dif\gamma_r^{\textrm{R}} = B_r\big( \Psi.\Gamma, \sum_{r'\neq r}\dif\gamma_{r'}^{\textrm{T}}, G \big)$
\ENDFOR
\FOR{$r=a,p,t$}
\State $\dif\gamma_r^{\textrm{T}} = 
	\dif\gamma_r^{\textrm{L}} + \dif\gamma_r^{\textrm{C}} + \dif\gamma_r^{\textrm{R}}$
\State $\dif\gamma_r \leftarrow \dif\gamma_r + \dif\gamma_r^{\textrm{T}}$
\ENDFOR
\State $\dif\gamma_{\bar{t}}^{\textrm{C}} \leftarrow 
	\dif\gamma_{\bar{t}}^{\textrm{C}} + \dif\gamma_a^{\textrm{C}}+ \dif\gamma_p^{\textrm{C}}$

\IF{$\textrm{max}_r \{ ||\dif\gamma_r^{\textrm{T}}||/||\dif\gamma_r|| \} < \epsilon$}
\State \textbf{break}
\ENDIF

\ENDFOR

\Statex /* \textit{jump to line 41 for $\ell_f$-loop fRG without corrections to the self-energy flow} */

\State $\dif\Sigma_{\bar{t}} = 
	L( \dif\gamma_{\bar{t}}^{\textrm{C}}, G) $
\State $\dif\Sigma_{t} = 
	L( \Psi.\Gamma, G \cdot \dif\Sigma_{\bar{t}} \cdot G)$
\State $\dif\Psi.\Sigma = \dif\Sigma_{\textrm{std}} + \dif\Sigma_{\bar{t}} + \dif\Sigma_{t}$

\IF{$|| S+G\cdot\dif\Psi.\Sigma\cdot G - \dif G|| / ||\dif G|| < \epsilon$}
\State \textbf{break}
\ENDIF

\ENDFOR

\State $\dif\Psi.\Gamma = \sum_r \dif\gamma_r$
\State \textbf{return} $\dif\Psi$

\end{algorithmic}
\label{alg}
\caption{Pseudocode for computing the r.h.s.\ of the mfRG flow 
for a given state of the system $\Psi$
(containing $\Psi.\Sigma$ and $\Psi.\Gamma$)
and a scale parameter $\Lambda$.}
\end{alg-table}
In this section, we present a pseudocode for the approximate
solution strategy of the mfRG flow explained in \SR{sec:mfRG_comp}.
Generally, an ordinary differential equation (ODE)
is of the form $\partial_{\Lambda} \Psi(\Lambda) = f(\Lambda, \Psi)$,
and numerous numerical ODE solvers are available.
The only input required for such an ODE solver,
apart from stating the initial condition $\Psi(\Lambda_i) = \Psi_i$
and the extremal points $\Lambda_i$, $\Lambda_f$,
is an implementation of the function $f(\Lambda, \Psi)$.
In the case of mfRG, $\Psi$---describing the state of the
physical system at a specified value of the flow 
parameter $\Lambda$---is a vector that contains
the self-energy (say, $\Psi.\Sigma$) and 
the vertex (say, $\Psi.\Gamma$)
for all configurations of quantum numbers
(e.g., Matsubara frequency, momenta, and spin).
In order to use an ODE solver to compute the mfRG flow, 
we only need to specify a way to compute $f(\Lambda, \Psi)$.
This is provided by \AlR{alg}, 
written in pseudocode.
\AlR{alg} makes use of functions outlined in the main text,
for which we also include dependencies that have been
suppressed earlier. This applies to the single-scale propagator
$S$ [\ER{eq:singlescale}] in line 1, the Dyson equation for $G$ [\ER{eq:dyson}]
in line 2, the differentiated bubble $\dot{B}$ [\ER{eq:gdotbubble}] in line 8,
and the bubble $B$ [\ER{eq:bubbles}], which is used several times.
For a good numerical performance, an efficient implementation
of the bubble functions appearing in \AlR{alg}
using vertex symmetries and high-frequency asymptotics
is crucial \cite{Wentzell2016, SuppKugler2017}.
The algorithm has a few external parameters:
$\ell_f$ (line 19) denotes the maximal loop order, and
$it_f$ (line 5) the number of iterations that improve
the accuracy of the flow by four orders of the interaction
with each step (cf.~\SR{sec:mfRG_comp}).
These parameters can also be used dynamically via the 
break conditions of the loops depending on the tolerance
$\epsilon$ (lines 30, 37). 
Note that, typically, one also specifies a tolerance 
for the numerical ODE solver, say $\epsilon_{\textrm{ODE}}$.
If $\epsilon$ is chosen in accordance with $\epsilon_{\textrm{ODE}}$
and the number of loops ($\ell_f$) or iterations ($it_f$)
is not fixed a priori, this algorithm yields a 
solution of the \textit{full} mfRG flow and thus a
full summation of \textit{all} parquet diagrams---to the
specified numerical accuracy.
The straightforward implementation as given by the pseudocode
in \AlR{alg} demonstrates the feasibility of the mfRG 
flow for almost any fRG application. 
\section{Multiloop flow beyond the parquet approximation}
\label{appendix2}
The mfRG flow as described so far achieves a full summation
of all parquet diagrams of the vertex and self-energy.
The first deviations from the exact quantities,
i.e., the first nonparquet diagrams, occur at fourth order for
the vertex---these are the envelope
vertices, such as the one shown in \FR{fig:sd}(a)---%
and, as follows by use of the SDE \ERn{eq:SchwingerDyson},
at fifth order for the self-energy.
One can in principle add terms to the mfRG flow equations
that go beyond the parquet approximation. 
The flow equation of $\Gamma$ then also needs to generate
differentiated diagrams of envelope vertices.
This is achieved by adding the differentiated envelope vertices,
i.e., all envelope diagrams of $\Gamma$ 
with one $G$ line replaced by $\partial_{\Lambda} G$
at all possible positions,
to the flow equation. Subsequently,
one performs the replacement $\Gamma^0 \to \Gamma$
to generate contributions at all interaction orders.
(Note that the mfRG
corrections of the self-energy flow 
have to be changed accordingly.)
However, such contributions to the vertex flow are---by the very fact
that they are of nonparquet type---not of an iterative one-loop structure anymore
[i.e., their evaluation requires the computation of
two or more (nested) integrals]
and are thus computationally unfavorable.
Another possibility to obtain nonparquet diagrams from mfRG
is to keep the flow equations unchanged and modify the initial condition.
One can then add scale-independent envelope vertices,
i.e., envelope vertices computed in the final theory (at $\Lambda_f$)
with some approximation of the self-energy, to the initial
condition of the vertex: 
$\Gamma_{\Lambda_i}^{\vphantom{a}} = \Gamma^0 + \Gamma^{\textrm{envelope}}_{\Lambda_f}$.
(Hence, $\Gamma^{\textrm{envelope}}$
must be computed only once.)
This yields contributions to the flow that are not
actually differentiated diagrams at a given scale $\Lambda$.
Nevertheless, the initial vertex $\Gamma_{\Lambda_i}$
constitutes a new totally irreducible building block in the mfRG flow.
After completion of the flow, one obtains a summation
of all ``parquet'' diagrams with the totally
irreducible vertex $R = \Gamma_{\Lambda_i}$
instead of $R = \Gamma^0$;
i.e., one obtains vertex and self-energy at one level
beyond the parquet approximation [cf.~\ER{eq:parquet}].
Such results deviate from the exact quantities starting at
fifth and sixth order for $\Gamma$ and $\Sigma$, respectively.
This scheme of adding nonparquet contributions can also be iterated and
used with expressions for $R = \Gamma_{\Lambda_i}$ of even higher order.
However, it appears rather tedious
and is more in the spirit of an iterative solution of the
parquet equations than of an actual fRG flow.
\bibliographystyle{apsrev4-1}
\bibliography{references}
\end{document}